\documentclass[manuscript]{aastex63}
\usepackage{amsmath}

\shorttitle{Data-driven simulation of the filament eruption}

\begin{document}

\title{Thermodynamic and Magnetic Topology Evolution of the X1.0 Flare on 2021 October 28 Simulated by a Data-driven Radiative Magnetohydrodynamic Model}

\correspondingauthor{P. F. Chen and Y. Guo}
\email{chenpf@nju.edu.cn, guoyang@nju.edu.cn}

\author[0000-0002-4205-5566]{J. H. Guo}
\affiliation{School of Astronomy and Space Science and Key Laboratory of Modern Astronomy and Astrophysics, Nanjing University, Nanjing 210023, China}
\affiliation{Centre for mathematical Plasma Astrophysics, Department of Mathematics, KU Leuven, Celestijnenlaan 200B, B-3001 Leuven, Belgium}

\author[0000-0002-9908-291X]{Y. W. Ni}
\affiliation{School of Astronomy and Space Science and Key Laboratory of Modern Astronomy and Astrophysics, Nanjing University, Nanjing 210023, China}

\author[0000-0001-5483-6047]{Z. Zhong}
\affiliation{Center for Integrated Research on Space Science, Astronomy, and Physics, Institute of Frontier and Interdisciplinary Science, Shandong University, Qingdao 266237, China}

\author[0000-0002-9293-8439]{Y. Guo}
\affiliation{School of Astronomy and Space Science and Key Laboratory of Modern Astronomy and Astrophysics, Nanjing University, Nanjing 210023, China}

\author[0000-0002-7153-4304]{C. Xia}
\affiliation{School of Physics and Astronomy, Yunnan University, Kunming 650050, China}

\author[0000-0001-6024-8399]{H. T. Li}
\affiliation{School of Astronomy and Space Science and Key Laboratory of Modern Astronomy and Astrophysics, Nanjing University, Nanjing 210023, China}

\author[0000-0002-1743-0651]{S. Poedts}
\affiliation{Centre for mathematical Plasma Astrophysics, Department of Mathematics, KU Leuven, Celestijnenlaan 200B, B-3001 Leuven, Belgium}
\affiliation{Institute of Physics, University of Maria Curie-Skłodowska, Pl.\ M.\ Curie-Sk{\l}odowska 5, 20-031 Lublin, Poland}

\author[0000-0003-3364-9183]{B. Schmieder}
\affiliation{Centre for mathematical Plasma Astrophysics, Department of Mathematics, KU Leuven, Celestijnenlaan 200B, B-3001 Leuven, Belgium}
\affiliation{LESIA, Observatoire de Paris, CNRS, UPMC, Universit\'{e} Paris Diderot, 5 place Jules Janssen, 92190 Meudon, France}

\author[0000-0002-7289-642X]{P. F. Chen}
\affiliation{School of Astronomy and Space Science and Key Laboratory of Modern Astronomy and Astrophysics, Nanjing University, Nanjing 210023, China}

\begin{abstract}
Solar filament eruptions, flares and coronal mass ejections (CMEs) are manifestations of  drastic release of energy in the magnetic field, which are related to many eruptive phenomena from the Earth magnetosphere to black hole accretion disks. With the availability of high-resolution magnetograms on the solar surface, observational data-based modelling is a promising way to quantitatively study the underlying physical mechanisms behind observations. By incorporating thermal conduction and radiation losses in the energy equation, we develop a new data-driven radiative magnetohydrodynamic (MHD) model, which has the capability to capture the thermodynamic evolution compared to our previous zero-$\beta$ model. Our numerical results reproduce major observational characteristics of the X1.0 flare on 2021 October 28 in NOAA active region (AR) 12887, including the morphology of the eruption, kinematic of flare ribbons, extreme-ultraviolet (EUV) radiations, and two components of the EUV waves predicted by the magnetic stretching model, i.e., a fast-mode shock wave and a slower apparent wave due to successive stretching of magnetic field lines. Moreover, some intriguing phenomena are revealed in the simulation. We find that flare ribbons separate initially and ultimately stop at the outer stationary quasi-separatrix layers (QSLs). Such outer QSLs correspond to the border of the filament channel and determine the final positions of flare ribbons, which can be used to predict the size and the lifetime of a flare before it occurs. In addition, the side view of the synthesized EUV and white-light images exhibit typical three-part structures of CMEs, where the bright leading front is roughly cospatial with the non-wave component of the EUV wave, reinforcing the magnetic stretching model for the slow component of EUV waves.

\end{abstract}

\keywords{Radiative magnetohydrodynamics (2009); Solar coronal mass ejections (310); Solar magnetic fields (1503); Solar prominences (1519); Solar flares (1496)}

\section{Introduction} \label{sec:intro}

Solar eruptions, such as filament eruptions, solar flares and coronal mass ejections \citep[CMEs;][]{Chen2011, Webb2012, Toriumi2019}, correspond to the sudden and explosive release of magnetic energy in the corona via magnetic reconnection, which are usually associated with particle acceleration, and various types of waves, constituting the prominent features of many eruptive astrophysical processes. In general, the core field of the pre-eruption system is either a twisted magnetic flux rope or a sheared arcade \citep{Ouyang2015}. After being triggered, the eruptions usually excite waves and wave-like perturbations, including coronal EUV waves and chromospheric Moreton waves \citep{Vrank2008, Warmuth2015, Chen2016}. When a CME propagates in the heliosphere, it is termed interplanetary coronal mass ejection (ICME) or magnetic cloud \citep[MC;][]{Burlaga1981, Tsurutani2022}. The magnetized ejecta following the enhanced flaring electromagnetic emissions, along with the associated energetic particles,  are the main disturbing sources of the heliospheric environment. Therefore, studies of solar eruptions not only deepen our understanding on basic astrophysical processes, but also improve the capacity of predicting adverse space weather events.

Although these eruptive phenomena appear to be significantly different, they can be linked by the standard CSHKP flare model \citep{Carmichael1964, Sturrock1966, Hirayama1974, Kopp1976}. In this model, the core magnetic field, either a flux rope or a sheared arcade, is triggered to ascend due to an ideal MHD instability or magnetic reconnection \citep{Chen2011}, inducing the formation of a current sheet below the core field. The magnetic reconnection inside the current sheet produces downward outflows and flaring loops below the reconnection site and upward flow with concave-upward magnetic field above the reconnection site. The ``slingshot'' effect of the magnetic field in the upward flow pushes the core field, which might host a filament, to move upward and ultimately evolve into a CME \citep{Jiang2021}. Furthermore, the reconnection of magnetic field converts the magnetic energy to thermal and kinetic energies in a bulk of materials, in addition to a large number of energetic particles traveling along magnetic field lines. Energetic particles and heat conduction propagating downward heat the plasma in the lower atmosphere to form flare ribbons, and cause chromospheric evaporation to form flare loops simultaneously \citep{Hudson2011}. As the flux rope keeps rising, more field lines participate in the reconnection, and newly-formed larger loops overlie on the previous ones, manifested as the continuous expansion of flare loops and separation of two main ribbons \citep{Petrasso1979}. Albeit the key processes are described well by the above physical paradigm, some issues have not yet been well addressed. For example, where do the separating flare ribbons stop \citep{Chen2012}? Besides, a lot of observations showed that a CME generally presents a three-part structure, that is, bright core, dark cavity and bright leading front \citep{Chen2011}. However, it is still not clear how the CME leading fronts are formed. In contrast to the general belief that they are due to plasma pileup ahead of the eruption, \citet{Chen2009} proposed that the CME leading fronts correspond to the density enhancement due to successive magnetic field line stretching. As for the CME cores, \citet{Song2022} suggested that both the warm channel and prominence can evolve into the CME core. 

Apart from these, the classical 2-dimensional (2D) model encounters many difficulties in quantitatively explaining some 3-dimensional (3D) features in observations. First, in the 2.5D model, there is strict distinction between the toroidal and poloidal magnetic fields, and the toroidal flux keeps constant as in the pre-eruptive structure. However, in 3D cases, there is no strict distinction between toroidal and poloidal fields, and there exists mutual conversion, for example, observations revealed that reconnection can contribute a substantial amount of toroidal flux to the flux rope \citep{Qiu2007, Wangws2017, Xing2020a}. Second, the footpoints of the flux rope might drift significantly in real eruptions \citep{Aulanier2019, Zeman2019, Xing2020a}, which is hard to be explained by 2D models. To address these issues, \citet{Aulanier2019} analyzed the results of a 3D flare model based on a zero-$\beta$ MHD simulation, which contains three types of reconnection geometries: the reconnection in the overlying arcades (aa-rf reconnection), the reconnection between ambient sheared arcades and the flux rope (ar-rf reconnection), and the reconnection between two flux-rope legs (rr-rf reconnection). This 3D flare model sets the stage for explaining complex observations \citep{Zeman2019, Xing2020a}, which is worthy of validation and further refinement by comparing simulations and observations.

In recent years, numerical models based directly on observational data are drawing ever more attention. In these models, the observational data in the photosphere, such as the magnetic field \citep{Jiang2016}, velocity \citep{Hayashi2019, Kaneko2021, Jiang2021f}, the combination of velocity and magnetic field \citep{Guo2019b} and electric field \citep{Cheung2012, Hayashi2018, Pomoell2019, Fisher2020}, is taken as the input to drive the evolution of the coronal magnetic field and plasma. This kind of data-driven or data-constrained models can be directly compared with the multi-wavelength observations, presenting great power in quantitatively explaining the underlying physical mechanisms behind observations \citep{Jiang2022}. For example, \citet{Jiang2018a} reproduced the corona loops, filaments, and flare ribbons in the largest solar flare in solar cycle 24. \citet{Guo2019a} demonstrated that the torus instability \citep{Kliem2006, Aulanier2010} plays a pivot role in the triggering of a flux rope eruption. \citet{Zhong2021} found that the Lorentz force component resulting from the non-axisymmetry of the flux rope can constrain the eruption. However, it is noticed that, to reduce computing expenses,  many authors adopted the time-dependent magnetofrictional model \citep{Cheung2012, Cheung2015, Pomoell2019, Kilpua2021, Lumme2022}, zero-$\beta$ model \citep{Kliem2013, Inoue2018, Guo2019a, Zhong2021, Guoy2021}, or isothermal MHD model \citep{Jiang2013, Jiang2016a, Jiang2016b, Jiang2018a, Jiang2018b}, in which the thermal properties of the plasma are discarded or simplified. This means that the above models are incapable of synthesizing the radiations from the density and temperature self-consistently, and cannot be directly compared with EUV observations. Recently, some data-based MHD models that take into account the non-adiabatic effects have been performed to reproduce the CME \citep{Fan2022}. In short, to better understand the nature of some emission structures (such as filaments, EUV waves and coronal loops), we need to develop data-driven radiative MHD models. 

Here, we develop and perform a data-driven radiative MHD simulation of the X1.0 flare on 2021 October 28. The initial magnetic field is obtained with a non-linear force-free field (NLFFF) reconstruction. Thermal conduction and radiation losses are considered in the energy equation. We have two goals in this paper: the first is to reproduce most of the observational features, and the second is to study thermodynamics and the magnetic topology evolution during the eruption, especially the flare ribbons and CME leading front. The event overview and numerical methodology are described in Section~\ref{sec:method}. Numerical results are displayed in Section~\ref{sec:results}, which are followed by discussions in Section~\ref{sec:discussion}. Finally, we summarize our results and discoveries in Section~\ref{sec:summary}.

\section{Event overview and numerical setup} \label{sec:method}

\begin{figure}[ht!]
\centering
\includegraphics[scale=0.60]{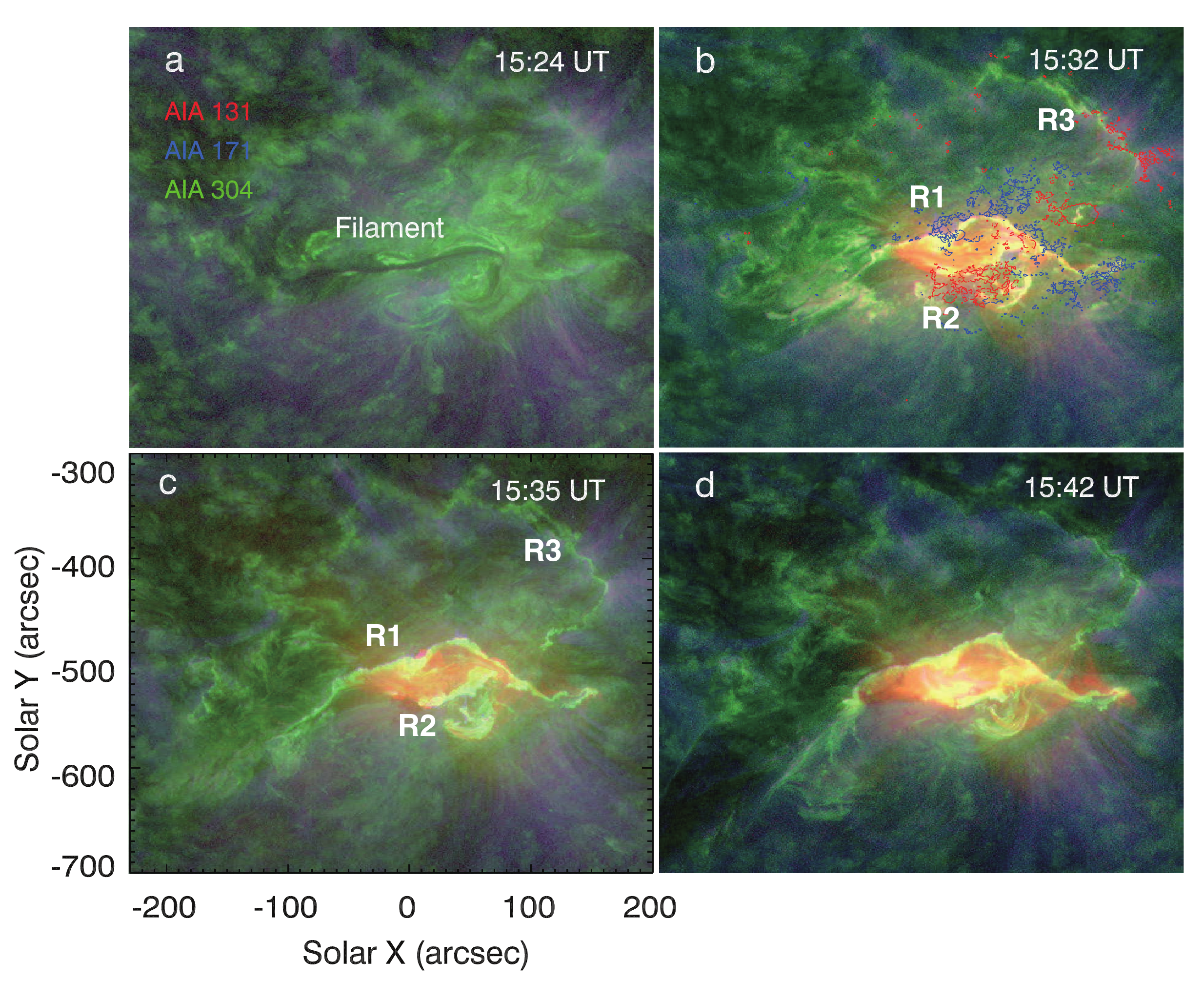}
\caption{Multi-wavelength composite images of the SDO/AIA at 15:24, 15:32, 15:35 and 15:42 UT on 2021 October 28, showing the evolution of the filament, flare ribbons and post-flare loops. The blue/red contour lines in panel (b) correspond to the positive/negative magnetic polarities.}
\label{figure1}
\end{figure}

The event we select is the second GOES X-class flare in solar cycle 25, hosted in NOAA AR 12887, occurring around 15:35 UT on 2021 October 28. This event has been reported by other papers, which focused on some observational features, such as the observations of EUV waves and the corresponding CME \citep{Hou2022, Devi2022}, CME three-part structure \citep{Devi2022}, eruption mechanism \citep{Yamasaki2022}, the Sun-as-a-star spectroscopic characteristics \citep{Xu2022}, solar energetic particles \citep{Lixl2022}, and its geomagnetic effects \citep{Papaioannou2022}. It would be of great interest to check in what degree such a geoeffective solar eruption can be reproduced with a data-driven radiative MHD model. In this paper, we mainly dig out the underlying physical properties, in particular the flare ribbons and CME leading front with the aid of numerical simulations. Figure~\ref{figure1} shows the composite images of the event at three EUV wavebands (131~\AA, 171~\AA, and 304~\AA) during eruption, which are provided by the Atmospheric Imaging Assembly \citep[AIA;][]{Lemen2012} on board the Solar Dynamics Observatory. There are two primary observational features. First, a long filament is located in this active region before the flare onset (Figure~\ref{figure1}a), implying that a pre-eruptive twisted/sheared magnetic structure exists before eruption. Second, the flare is composed of two main parallel ribbons (marked as R1 and R2) and a remote ribbon (R3). In particular, ribbon R1 is exceedingly long and narrow. Besides, one can see that only the eastern filament materials are still partly visible during eruption, reflecting that there might be deeper dips that are able to host filament materials for a longer time in the eastern portion of the flux rope than in the western portion. The above observations provide some constrained conditions for the following numerical setup of the data-driven model. 

For the numerical simulation, we consider a full MHD model including thermal conduction, empirical heating, and radiative cooling terms. The following are the governing equations:
\begin{eqnarray}
 && \frac{\partial \rho}{\partial t} +\nabla \cdot(\rho \boldsymbol{v})=0,\label{eq1}\\
 && \frac{\partial (\rho \boldsymbol{v})}{\partial t}+\nabla \cdot(\rho \boldsymbol{vv}+p_{_{tot}}\boldsymbol{I}-\frac{\ \boldsymbol{BB}}{\mu_{0}})=\rho \boldsymbol{g},\label{eq2}\\
 && \frac{\partial \boldsymbol{B}}{\partial t} + \nabla \cdot(\boldsymbol{vB-Bv})=0,\label{eq3}\\
 && \frac{\partial \varepsilon}{\partial t}+\nabla \cdot(\varepsilon \boldsymbol{v}+p_{_{tot}}\boldsymbol{v}-\frac{\boldsymbol{BB}}{\mu_{0}}\cdot \boldsymbol{v}) =\rho \boldsymbol{g \cdot v}+H_{0}e^{-z/\lambda}-n_{\rm e}n_{\rm _{H}}\Lambda(T) \label{eq4} \\        
 &&+ \nabla \cdot(\boldsymbol{\kappa} \cdot \nabla T), \notag
\end{eqnarray}
where $p_{_{tot}} \equiv p + B^2 / (2\mu_{0})$ is the total pressure with the assumption of full ionization, $\varepsilon =\rho v^2/2+p/(\gamma -1)+ B^2 / (2\mu_{0})$ is the total energy density, $\boldsymbol{g}=-g_{\odot}r_{\odot}^2/(r_{\odot}+z)^2\boldsymbol{e_{z}}$ is the gravitational acceleration, $g_{\odot}= \rm 274\ m\ s^{-2}$ is the gravitational acceleration at the solar surface, $r_{\odot}$ is the solar radius, $\boldsymbol{\kappa}=\kappa_{\parallel}\boldsymbol{\hat{b}\hat{b}}$ is field-aligned thermal conduction, $\kappa_{\parallel} =10^{-6}\ T^{\frac{5}{2}}\ \rm erg\ cm^{-1}\ s^{-1}\ K^{-1}$ is the Spitzer heat conductivity, $H_{0}=10^{-4} \rm \ erg \ cm^{-3} \ s^{-1}$, $n_{_{H}}$ is the number density of protons, $n_{e}$ is the number density of electrons, $\lambda=60\ \rm Mm$ \citep{Xia2016}, $\Lambda(T)$ is the radiative loss function, and the other parameters have their general meanings. Equation (\ref{eq4}) mainly takes into account an empirical heating term used to maintain the hot corona ($H_{0}e^{-z/\lambda}$), optically thin radiation ($n_{\rm e}n_{\rm _{H}}\Lambda(T)$), and thermal conduction ($\nabla \cdot(\boldsymbol{\kappa} \cdot \nabla T)$). The curve of the radiative loss function for our simulation is similar to that used in \citet{Zhao2017}, where the radiative cooling coefficient was calculated by \citet{Colgan2008}. It is worth noting that, the background heating function in this paper only varies with the height, which is principally used to compensate the coronal radiation losses, and widely used in simulations for filament formation and eruption \citep{Xia2016, Fan2017, Zhao2017, Fan2020}. In future works, we expect to consider a more realistic heating function inspired by the $\rm Alfv\acute{e}n$ turbulence dissipation \citep{Mok2016}. In addition, we use the transition region adaptive conduction (TRAC) method to correct the chromospheric evaporation, which can better capture the energy exchange between the transition region and corona with limited spatial resolution \citep{Johnston2019, Zhou2021}.

\begin{figure}[ht!]

\includegraphics[scale=0.4]{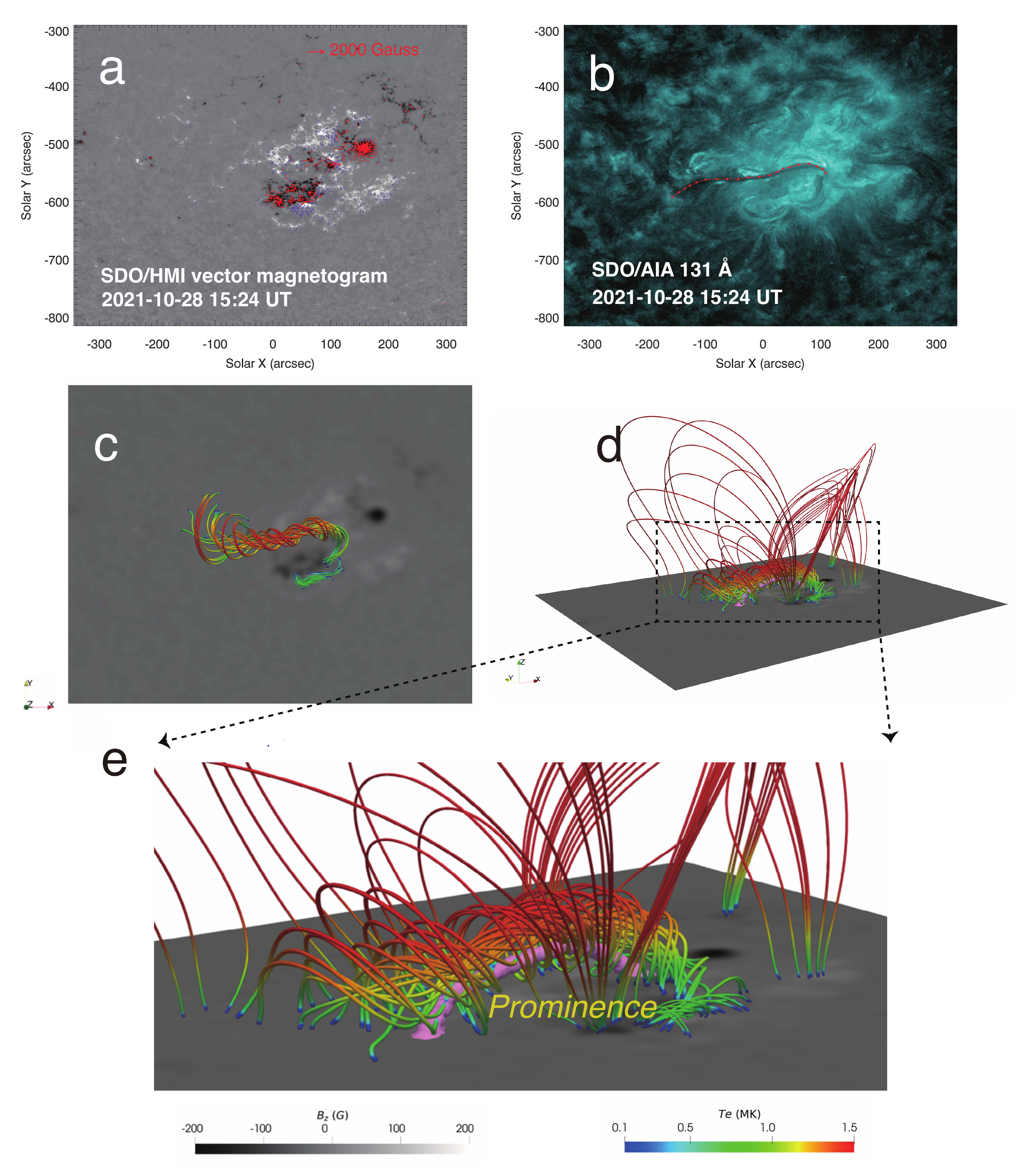}
\centering
\caption{(a) Remapped HMI vector magnetogram as viewed from the top. The $B_{z}$ is plotted in the background (gray scale), and the purple (red) arrows denote the horizontal magnetic fields with positive (negative) polarities. (b) AIA 131 \AA\ image at 15:24 UT on 2021 October 28, where the red dots outline the path of the inserted flux rope. (c) Selected core field lines of the NLFFF model viewed from the top,  colored in temperature. The temperature is obtained from the inserted filament and hydrostatic model. (d) Side view of the extrapolated field lines. (e) The constructed flux rope and inserting prominence with a zoomed-in view of the black rectangle in panel d, where the pink isosurface represents the prominence material where the temperature is lower than 20000 K.}
\label{figure2}
\end{figure}

The initial magnetic field $\boldsymbol{B}$ for the simulation is provided by the NLFFF model constructed by the Regularized Biot-Savart Laws \citep[RBSL;][]{Titov2018} and magneto-frictional \citep[MF;][]{Guo2016a, Guo2016b} method. The RBSL model is utilized to provide a flux rope consistent with the observational filament, and the MF method is to relax the magnetic field to a force-free state. The bottom boundary for the NLFFF model is provided by the corrected vector magnetic field in the photosphere at 15:24 UT observed by SDO/HMI \citep{Scherrer2012, Schou2012, Hoeksema2014}, as shown in Figure~\ref{figure2}a. To be specific, the pretreatment for the vector magnetogram contains two steps: to correct the projection effects resulting from the spherical shape of the solar surface \citep{Guo2017b}, and to remove the Lorentz force and torque in order to satisfy the force-free assumption \citep{Wiegelmann2006}. As introduced by \citet{Titov2018}, an RBSL flux rope is controlled by four parameters, that is, flux-rope path, minor radius, $a$, toroidal flux, $F$, and electric current, $I$. The minor radius of the flux rope is taken as twice the filament width, approximately 14~Mm, as a filament might occupy only half of the radial extent of the flux rope \citep{Guo2022}. Regarding the 3D path of the flux rope, similar to our previous work \citep{Guo2021a}, we outline the projected path according to the AIA observations at first, as indicated by the red dots in Figure~\ref{figure2}b. Then, we derive the height distribution of the flux rope using the triangulation technique \citep{Thompson2009} combined with the data from the \emph{Solar Terrestrial Relations Observatory} (\emph{STEREO}). However, we do not adopt the filament height directly as the flux-rope height, for two reasons. First, the filament is generally located at the bottom of the flux rope so that the filament is likely to deviate from the flux-rope axis \citep{Guo2021b, Guo2022}. Second, the angle between SDO and STEREO is relatively small ($\sim 37^{\circ} $), so it is prone to bring large measurement errors. As a result, we merely use the information of the filament apex as the apex of the flux rope axis, and the heights of the rest points are fitted with the following formula \citep{Torok2010, Xu2020}:
\begin{eqnarray}
 && z(x)=  \begin{cases} \frac{x(2x_{\rm h}-x)}{x_{\rm h}^{2}}h \quad 0 \le x \le x_{\rm h},\\ \frac{(x-2x_{_{\rm h}}+1)(1-x)}{(1-x_{\rm h})^{2}}h \qquad x_{_{\rm h}} < x \le 1 \end{cases} \label{eq5}
 \end{eqnarray}
where $x=s/l$ is the normalized coordinates along the filament axis, $x_{_{\rm h}} \sim 0.2$ is the normalized position of the filament apex, $h_{\rm f}=19\;$Mm is the height of the filament apex, $s$ is the distance from the start point, $l$ is the projection length of the filament, $h=h_{\rm f}+a$ is the height of the flux-rope apex. For the toroidal flux, similar to our previous works \citep{Guo2019b, Guo2021a}, the reference flux is selected as the average of the unsigned flux of two footprints, namely, $F_{0}=6.19 \times 10^{20}\;$Mx. According to our experience and some tests, we select $3.5F_{0}$ as the final toroidal flux. Then, the electric current, $I$, is computed by Equation~(12) in \citet{Titov2018}. It is noted that the sign of the electric current determines the magnetic helicity of the flux rope. In this event, the filament is located in the southern hemisphere, implying that the chirality of the filament is most probably sinistral according to the hemispheric rule \citep{Ouyang2017}. Moreover, the axial magnetic field of the filament points to the left if we stand on the positive side of the polarity inversion line. These strongly suggest that the chirality of the filament is sinistral, corresponding to positive helicity of the flux rope and electric current. So far, we can construct the RBSL flux rope and embed it into the potential field extrapolated by the Green's function method \citep{Chiu1977}, and then relax the magnetic field to a force-free state with the MF method \citep{Guo2016a, Guo2016b}. After relaxation, the force-free metric is $\sigma_{J}=0.29$, and the divergence-free metric is $\langle |f_{i}|\rangle =1.0 \times 10^{-5}$ \citep[see][for details of the two metrics]{Guo2016b}. Figure~\ref{figure2}c shows the sheared and twisted field lines after relaxation. One can see that the reconstructed flux rope and sheared arcades resemble the observed filaments. 

To obtain a hydrostatic atmospheric model from the chromosphere to the corona, similar to \citet{Xia2016}, the temperature distribution is described by the following formula:
\begin{eqnarray}
 T(z)=  \begin{cases} T_{_{\rm ch}}+(T_{_{\rm co}}-T_{_{\rm ch}})(1+T_{_{\rm ch}}+{\rm tanh}(z-h_{_{\rm tr}}- 0.27)/w_{\rm tr})/2 \qquad  z<h_{_{tr}},\\ (7F_{\rm c}(z-h_{\rm tr})/(2\kappa)+T_{_{\rm tr}}^{7/2})^{2/7} \qquad z \geq h_{_{tr}} \end{cases} \label{eq6}
 \end{eqnarray}
where $T_{_{\rm ch}}=8000 \rm \ K$ is the chromospheric temperature, $T_{_{\rm co}}=1.5 \rm \ MK$ is the coronal temperature at the top, $h_{_{\rm tr}}=2 \rm \ Mm$ is the height of initial transition region, $w_{_{\rm tr}}=0.2 \ \rm Mm$ determines the thickness of initial transition region, and $F_{\rm c}=2 \ \times 10^{5}\ \rm erg \ cm^{-2}\ s^{-1}$ is the constant thermal conduction flux. Then, with the assumption of the hydrostatic atmosphere, the density distribution can be derived from the given number density at the bottom of $1.15 \times 10^{15}\rm \ cm^{-3}$. Finally, we need to insert a filament according to the observation, which can be done by increasing the density by about 50 times of magnitude but keeping the gas pressure unchanged \citep{Zhou2018}, where the prominence path is measured from observations. Figure~\ref{figure2}d shows the side view of the 3D magnetic field lines, temperature distributions and the inserting filament, one can see that the cold filament materials almost reside in the bottom of the flux rope, consistent with the simulated filament formed with the thermal instability or the thermal non-equilibrium process \citep{Xia2014, Guo2021b, Guo2022}.

Similar to our former works \citep{Guo2019a, Zhong2021}, the $v$--$B$ driven boundary is adopted at the inner ghost cells of the bottom boundary. Concretely, we implement a sequence of vector magnetograms and velocity fields derived from the Differential Affine Velocity Estimator for Vector Magnetograms \citep[DAVE4VM;][]{Schuck2008} at the inner ghost cells so that the bottom boundary is synchronous with the observation. To reduce the small-scale fluctuations resulting from the uncertainty in the input magnetograms and to better match the boundary for the initial NLFFF model, we also adopt the preprocessing technique \citep{Wiegelmann2006} to process the time series of the driven magnetograms, where the smoothing term in the iteration function can effectively decrease the fluctuations of input magnetograms and the derived velocity fields. At the outer ghost cells of the bottom boundary, the magnetic field is provided by the fourth-order zero-gradient extrapolation, namely, $u(x_{i})=-0.12u(x_{i+4})+0.64u(x_{i+3})-1.44u(x_{i+2})+1.92u(x_{i+1})$. Additionally, since the cadence of the observational data is 12 min, the missing data within each 12 min are calculated with linear interpolation in time. For the other five boundaries, the velocity and magnetic field are provided by the equivalent extrapolation, $u(x_{i+1})=u(x_{i})$, and the second-order zero-gradient extrapolation, $u(x_{i})=(-u(x_{i+2})+4u(x_{i+1}))/3$, respectively. Regarding the density and gas pressure, they are fixed to be the initial values at the bottom, flexible at the top according to gravitational stratification \citep{Xia2016}, and provided by the equivalent extrapolation at four side boundaries. It is noted that, to save computing time and reduce the numerical dissipation, following previous simulations \citep{Jiang2016b, Kaneko2021}, we reduce the magnetic field strength to 1/11 of the original observed data. Still, the maximum of the driving velocity ($\sim$ 1 km s$^{-1}$) is less than the average $\rm Alfv\acute{e}n$ velocity at the bottom ($\sim$10 km s$^{-1}$).

The 3D MHD equations in the local Cartesian coordinate system are numerically solved with the Message Passing Interface Adaptive Mesh Refinement Versatile Advection Code \citep[MPI-AMRVAC\footnote{http://amrvac.org},][]{Xia2018, Keppens2020, Keppens2023}. For the numerical scheme, we use a three-step Runge-Kutta time discretization, HLL Riemann solver with the Cada limiter. The computational domain is $[x_{min},x_{max}]\ \times [y_{min},y_{max}]\ \times [z_{min},z_{max}] = [-242.6, 227.9]\ \times [-185.8,185.8]\ \times[1.0,400.0]\;$Mm, resolved by a uniform grid with $330 \times 260 \times 400$ cells. The time range of the simulation is from 15:24 to 16:00 UT in the observation, covering the whole process of the eruption.

\section{Numerical Results} \label{sec:results}

\subsection{Global evolution and comparison with observations} \label{sec:gc}

\begin{figure}[ht!]
\centering
\includegraphics[scale=0.25]{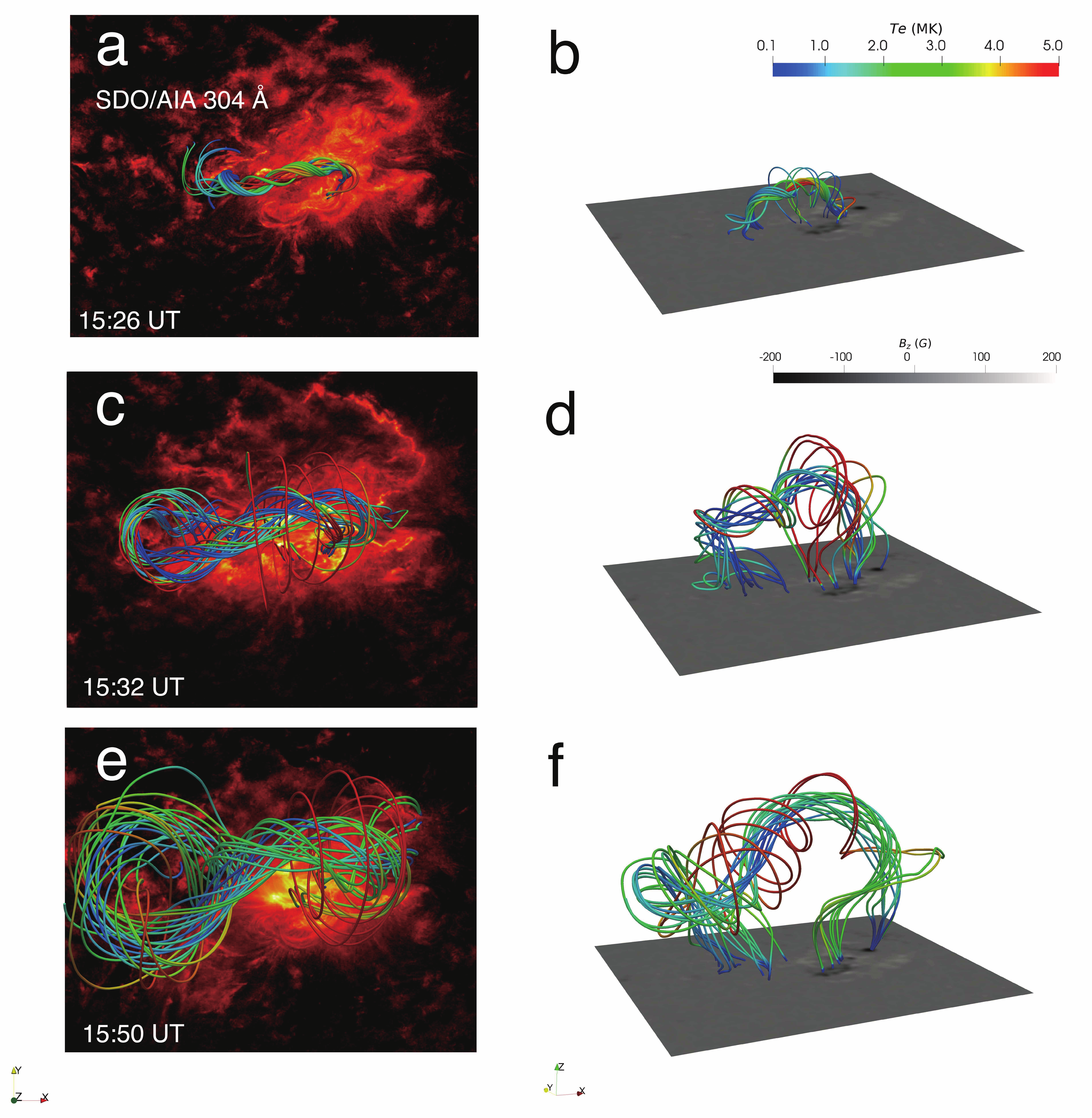}
\caption{Snapshots of the simulated magnetic field lines colored in temperature. The left column shows the evolution of the simulated flux ropes, which are overlaid on the 304 \AA\ images at (a) 15:26, (c) 15:32 and (e) 15:50 UT. The right column shows the evolution of the flux ropes from the side view at the same time as the left side.}
\label{figure3}
\end{figure}

The global evolution of the simulated flux rope during the eruption is shown in Figure~\ref{figure3}, which is overlaid on observed 304~\AA\ images in the left column and on the magnetograms in the right column. The morphology and erupting direction of the flux rope resemble those of the observed filament fairly well, indicating that the simulation almost reproduces the gross picture of the observation. Snapshots from the side views show drastic magnetic reconnection during the eruption, recognized by the hyperbolic flux tube \citep[HFT;][]{Titov2002} below the flux rope and the heating along the field lines (Figure~\ref{figure3}d). Moreover, the flux rope expands a lot and presents a stratified characteristic: the core field lines, which hold the cold prominence are weakly twisted, but the outer newly-formed field lines, which hold hot plasmas, are highly twisted, as shown in Figure~\ref{figure3}f. More intriguingly, the western footpoint of the flux rope is not fixed in the photosphere, which appears to drift westward. The drifting of the footpoint strongly implies that there should exist 3D reconnection geometry beyond the 2D CSHKP model \citep{Aulanier2019}.

\begin{figure}[ht!]
\centering
\includegraphics[scale=0.43]{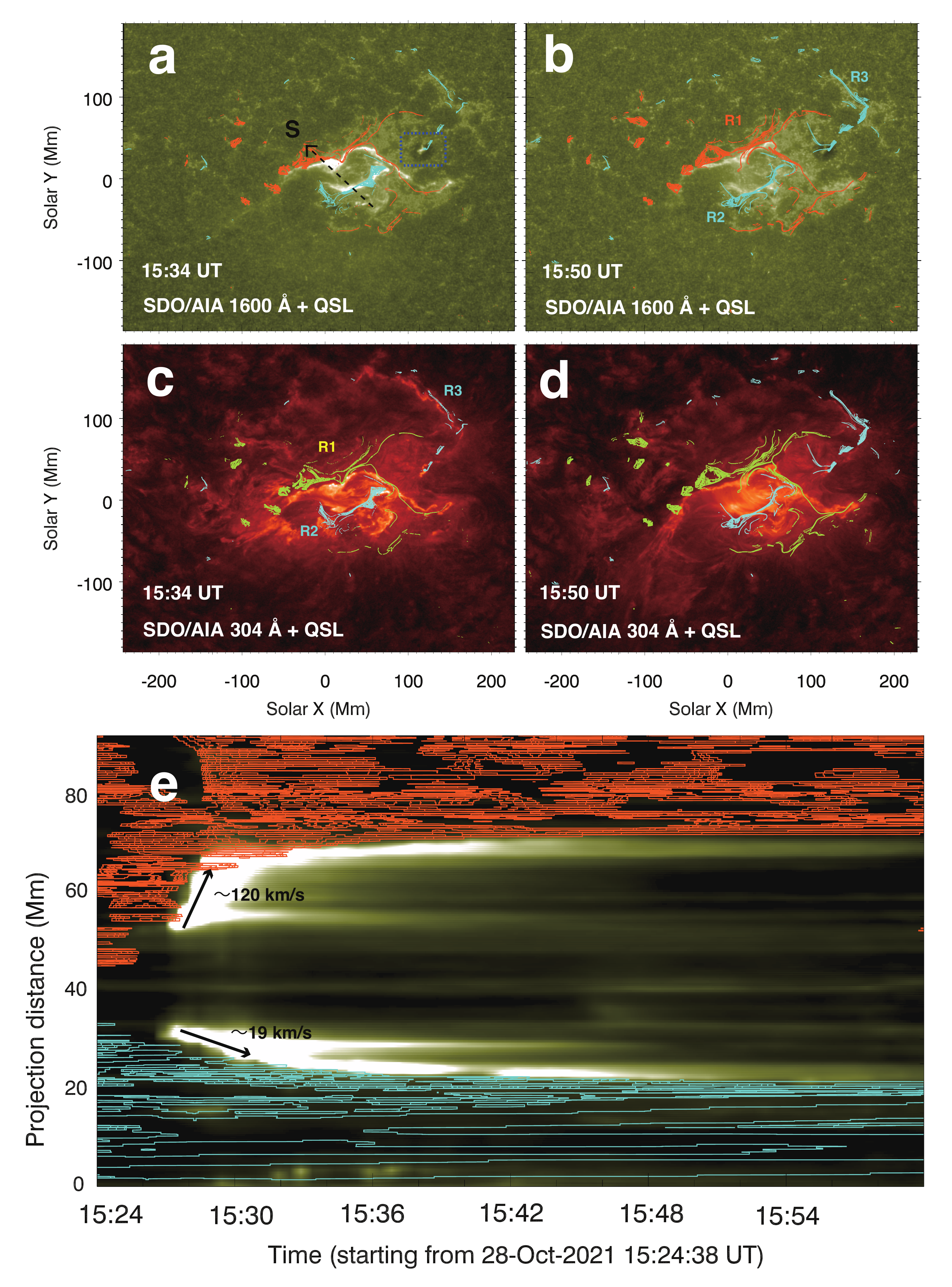}
\caption{Comparison between photospheric QSLs (lg $Q>$ 2) and the emission in AIA 1600 \AA\ (a--b) and 304 \AA\ (c--d). QSLs with the positive (negative) vertical magnetic component are colored in orange and chartreuse (cyan). The black dashed line in panel (a) illustrates a slice crossing the flare ribbons. The blue box shows the observation evidence of the 3D reconnection in Section \ref{re}. (e) Time-distance diagram of the QSLs and 1600 \AA\ image for slice S in panel (a).}
\label{figure4}
\end{figure}

 Quasi-separatrix layers (QSLs) are regions where magnetic connectivity changes drastically, and the squashing factor ($Q$) is larger than 2. They are the favorite places for magnetic reconnection \citep{Priest1995, demo96, Titov2002}. Former static NLFFF models and dynamic MHD simulations have shown that flare ribbons are almost cospatial with QSLs in the photosphere \citep{Demoulin1997, Janvier2013, Dudic2014, Yang2015, Dalmasse2015, Zuccarello2017,Jiang2018a, Guo2019a, Zhong2021}. Accordingly, their spatial relationship is an effective touchstone to examine the rationality of the simulation. To this end, we calculate the QSLs at each time step in the simulation with an open-source code \citep{Liu2016, Zhang2022} and compare them with the AIA 1600 and 304 \AA\ observations, as shown in the top two panels of Figure~\ref{figure4}. Note that the observational images are de-projected to a top view. We find that the simulated QSLs coincide well with the flare ribbons, including the two main parallel ribbons (R1 and R2) and the remote ribbon (R3). As expected, both R1 and R2 show separation motions. To further study their spatiotemporal relationship, we select a slice along the separation direction of the two parallel ribbons (Figure~\ref{figure4}a). The time-distance diagram in Figure~\ref{figure4}e illustrates that flare ribbons and QSLs show analogous kinematic characteristics: both the QSLs and the cospatial ribbons separate quickly and the flare ribbons eventually stop. The difference between the two ribbons is also remarkable. Ribbon R1 initially moves away from the polarity inversion line with a speed of $\sim$120 km s$^{-1}$, and then decelerates drastically. In contrast, ribbon R2 initially moves away from the polarity inversion line with a speed of $\sim$19 km s$^{-1}$, and then decelerates gradually. The comparability between the simulated QSLs and the observed flare ribbons strongly indicates that our numerical model reproduces the evolution of the magnetic topology in the observations very well.

\begin{figure}[ht!]
\centering
\includegraphics[scale=0.8]{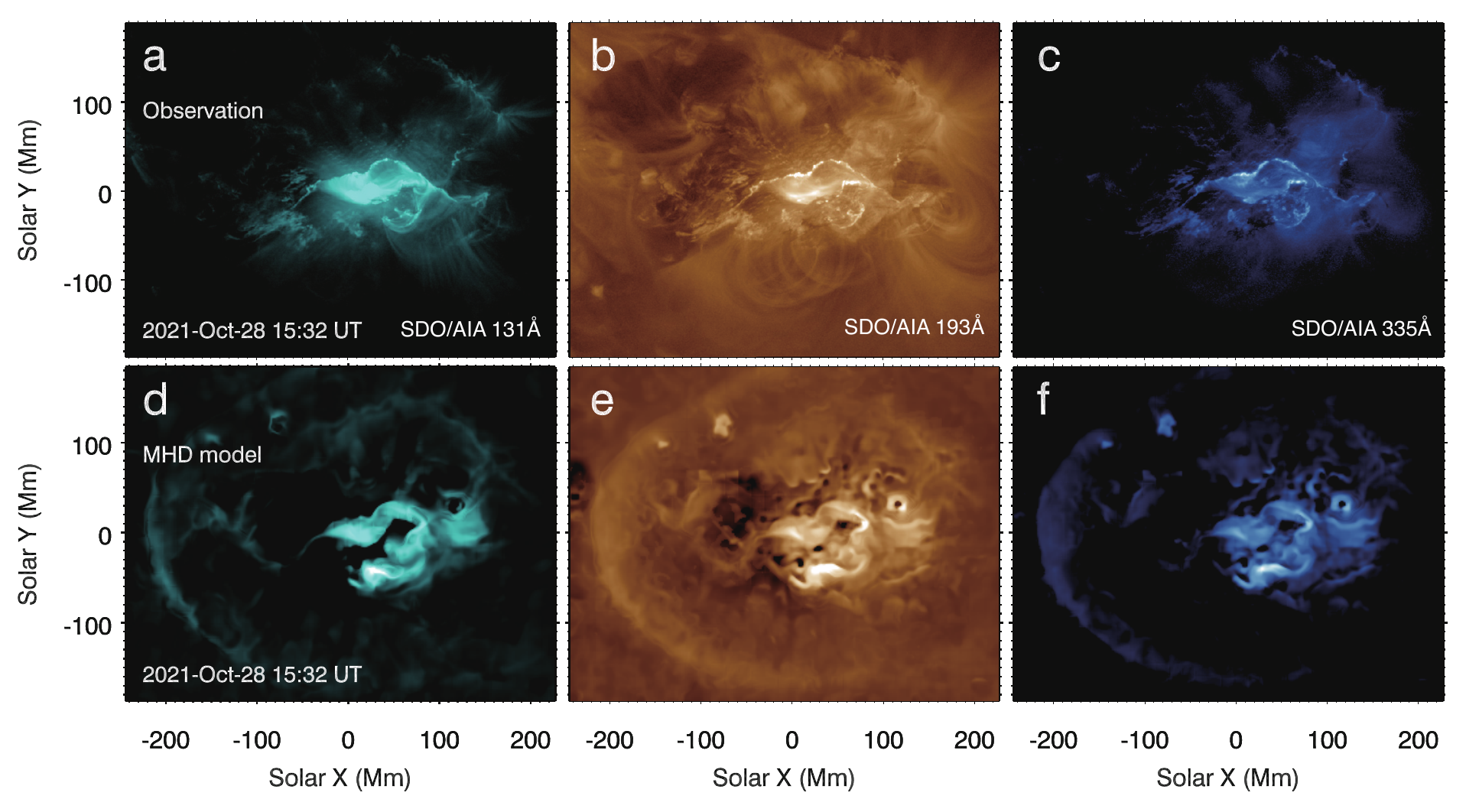}
\caption{Comparisons between AIA observations (a--c) and synthesized EUV radiative images (d--f) observed from the top at 15:32 UT. Panels (a) and (d) show the 131 \AA\ channel images, panels (b) and (e) show the 193 \AA\ channel images, and panels (c) and (f) show the 335 \AA\ channel images.}
\label{figure5}
\end{figure}

\begin{figure}[ht!]
\centering
\includegraphics[scale=0.8]{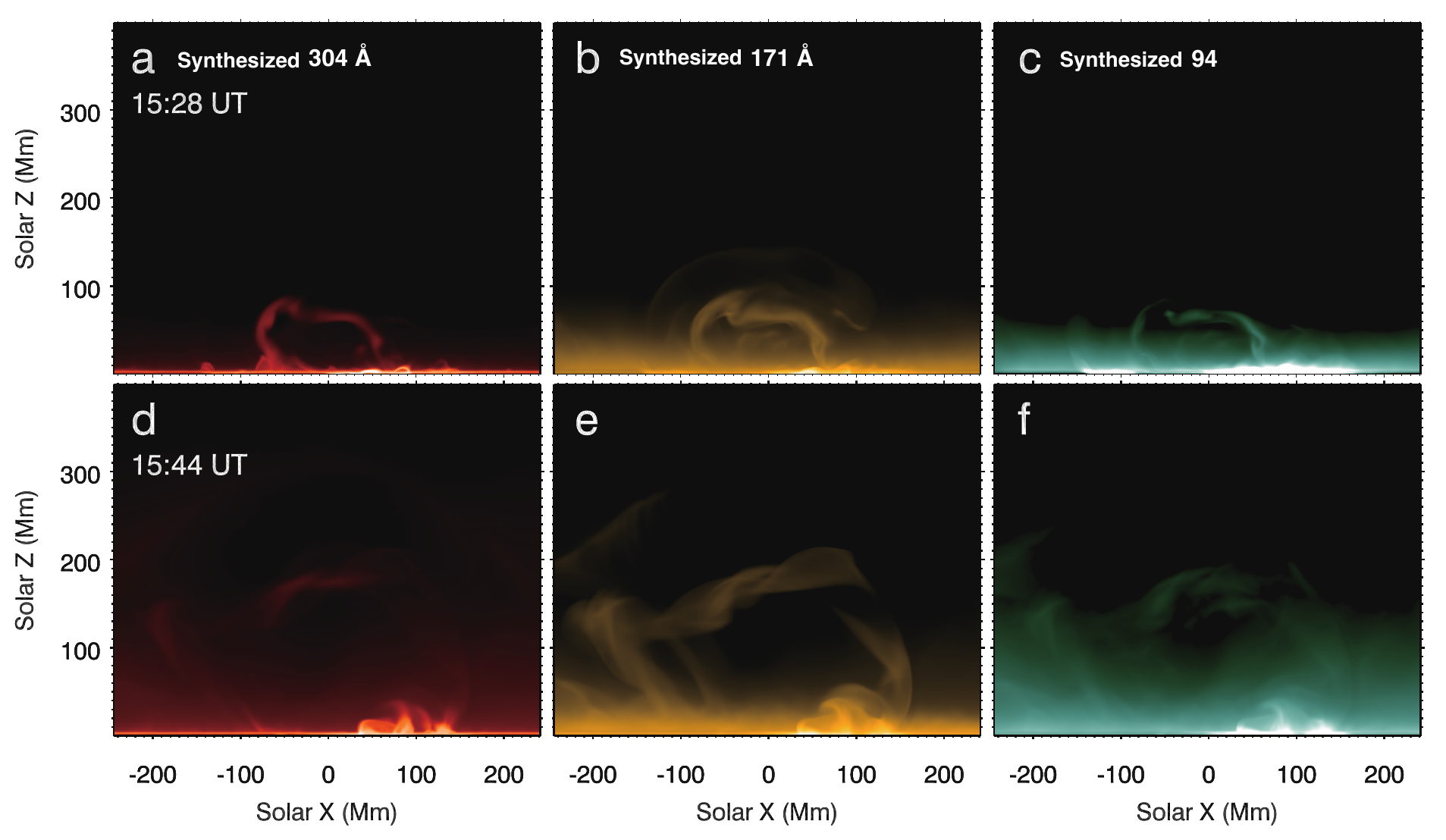}
\caption{Side views of the synthesized EUV radiations at 15:28 (a--c) and 15:44 UT (d--f). Panels from the left to right show the 304 \AA\ (a,d), 171 \AA\ (b,e) and 94 \AA\ (c,f) images, respectively.}
\label{figure6}
\end{figure}

\begin{figure}[ht!]
\centering
\includegraphics[scale=1.1]{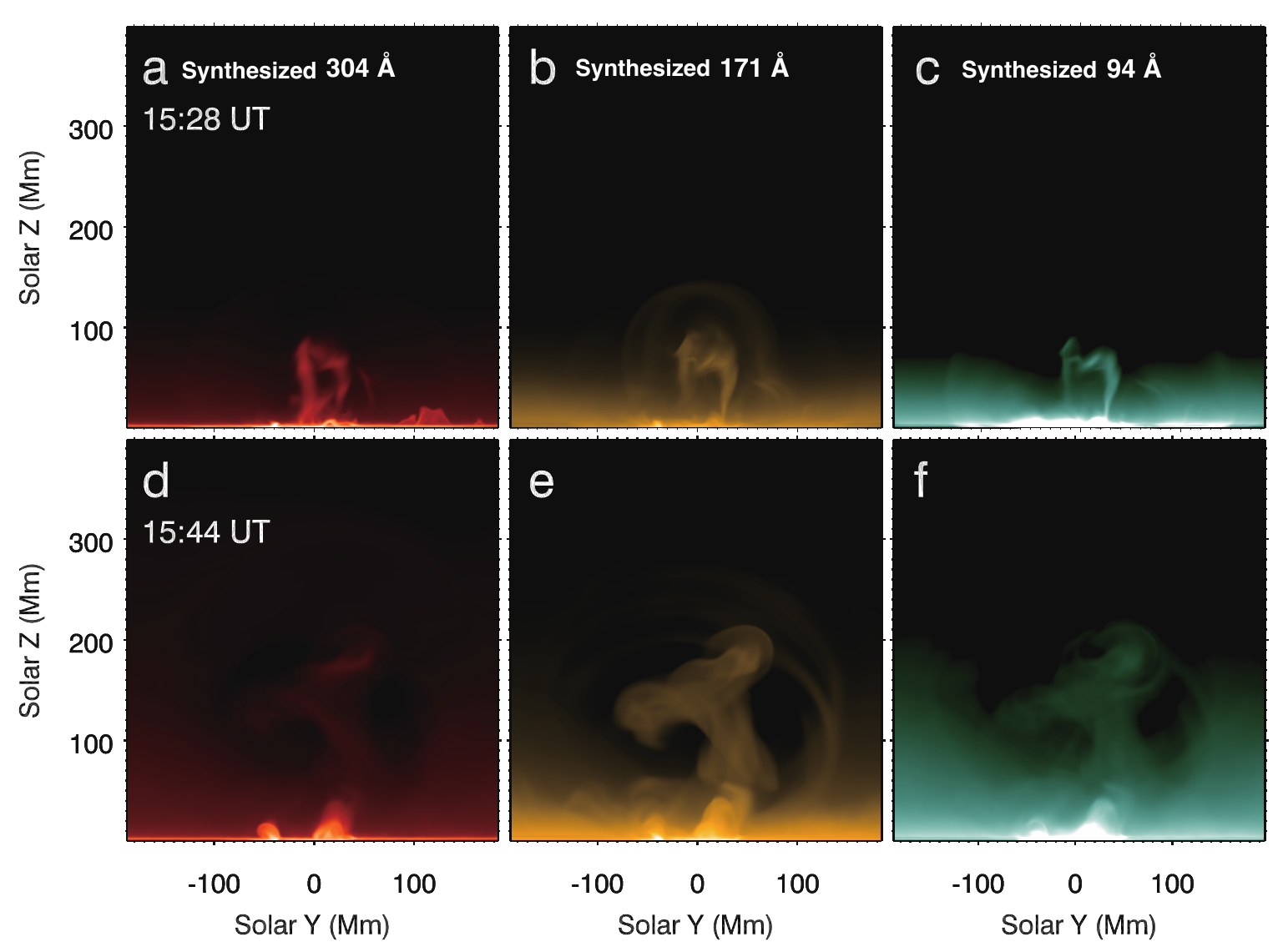}
\caption{Same as the Figure~\ref{figure6} but for the end views.}
\label{figure7}
\end{figure}

In addition to the evolution of the coronal magnetic field, our numerical model takes thermal processes into account. Consequently, we have the capability to compute the synthesized EUV images from the simulated temperature and density, and compare them with the EUV observations directly. The optically thin emission in each cell is calculated with the following formula \citep{Chen2006}:
\begin{eqnarray}
&& I_{_{\lambda}}(x,y,z)=G_{_{\lambda}}(T)\ n_{e}^{2}(x,y,z)  \label{eq7}
\end{eqnarray}
where $G_{_{\lambda}}(T)$ is the instrumental response function for different EUV wavebands. The synthesized image can be obtained by the integration along the line of sight. Figure~\ref{figure5} displays the comparisons between our synthesized EUV images (integration along the $z$-axis) and AIA observations at different wavelengths. It is found that the synthesized images show an analogous morphology to the observations, in particular the locations and the shapes of ribbon R1 and remote ribbon R3. An oval-shaped enhancement in the synthesized EUV images in panels (d--f) corresponds to the fast-component EUV wave on the solar surface. A similar wavelike structure is also discernable in observations as indicated by a slightly larger oval-shaped structure in Figure 5b. It implies that our simulated EUV wave is slightly slower than in observations. Figures~\ref{figure6} and \ref{figure7} display the side and end views of the synthesized EUV images, which are obtained by the integration along the $y$- and $x$-axes, respectively. Many common observational phenomena are presented in the synthesized EUV images, for example, the helical and eruptive prominence (Figure~\ref{figure6}e and \ref{figure7}e), flare loops (Figures~\ref{figure6}d and \ref{figure7}d), cusp-like structure (Figures~\ref{figure6}f and \ref{figure7}f) and CME three-part structure (Figure~\ref{figure7}b). As the eruption goes on, the filament observed in 304 \AA\ waveband becomes fainter gradually, which might be due to enhanced heating and the density decrease caused by the expansion of the flux rope. It is seen that the erupting flux rope can be seen in three wavebands, reflecting that it is a multithermal structure during the eruption. In addition, one can find that the majority of filament material stays in the eastern portion of the flux rope during the eruption (Figure~\ref{figure6}e), which is similar to the observations.

In conclusion, many observational features are reproduced well by our data-driven radiative model, in terms of the morphology of the eruption, dynamics of the magnetic topology and EUV radiations. These results indicate that the simulation is reasonable and provide a solid basis for the following quantitative analyses.

\subsection{Magnetic reconnection and topology evolution during the eruption}\label{re}

\begin{figure}[ht!]
\centering
\includegraphics[scale=0.7]{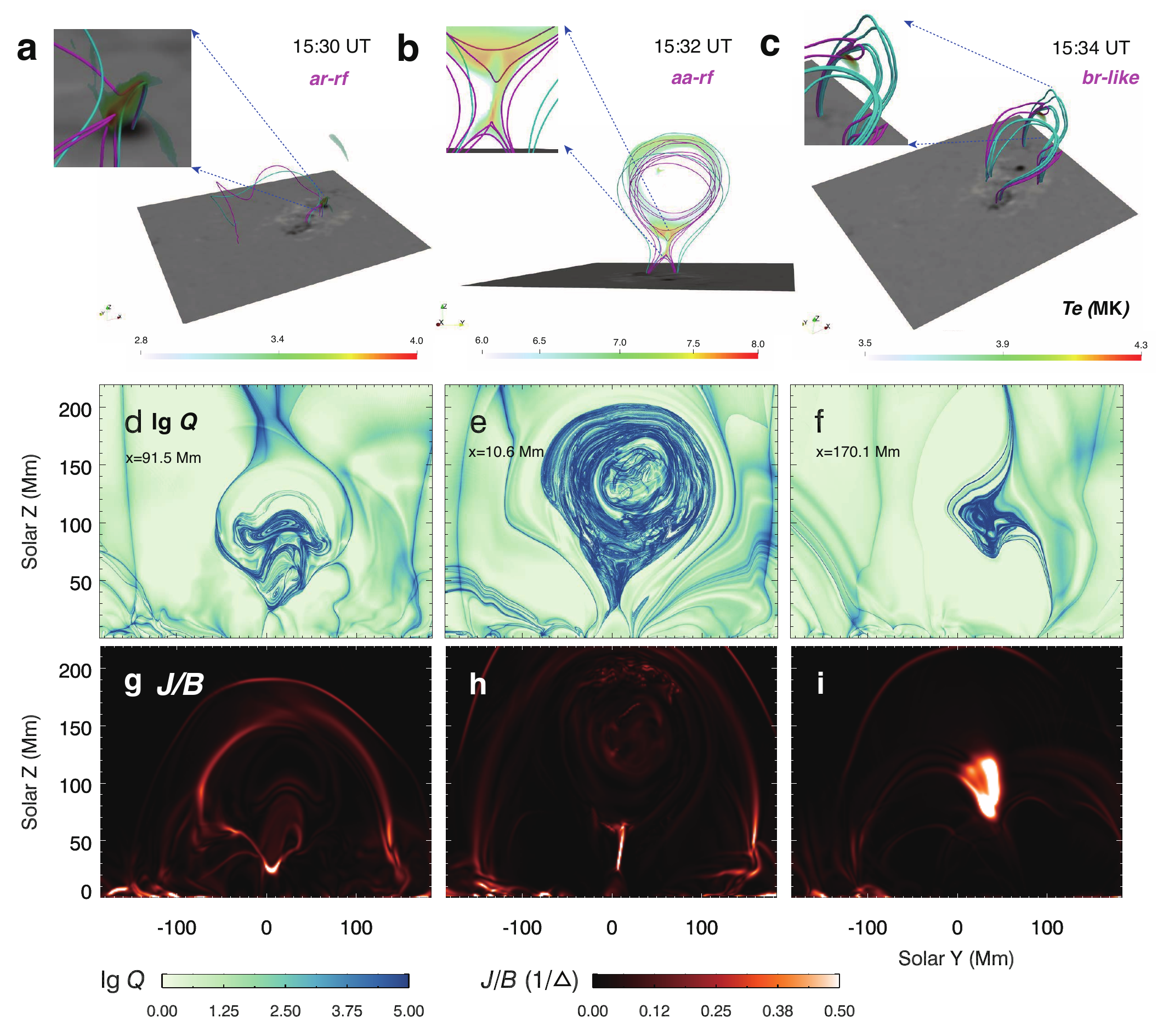}
\caption{Reconnection geometries recognized by the field-line connectivity (a--c), $Q$ distribution (d--f) and $J/B$ distribution (g--i). Columns from the left to right represent the ar-rf geometry at 15:30 UT, aa-rf geometry at 15:32 UT and breakout-like geometry at 15:34 UT, respectively. (a--c): Pre-reconnection (cyan) and post-reconnection (purple) field lines, where the transparent surfaces represent the plasma heated by the reconnection. (d--f): $Q$ distributions near the current sheet, which are correspond to three types of reconnection geometries displayed in panels (a--c). (g--i): $J/B$ distributions at the same plane of the second row.}
\label{figure8}
\end{figure}

Figure~\ref{figure8} displays the evolution of the distributions of the magnetic field lines, heating, QSLs, and $J/B$. It is noted that $J/B$ is a scalar metric to denote the variation of the magnetic field ($J/B \sim |\nabla \times B|/B$), which is widely used to locate the current sheet and magnetic reconnection \citep{Gibson2006, Fan2007, Jiang2016b, Jiang2018a, Jiang2018b, Jiang2021}. Three types of magnetic reconnection can be identified, as represented by the three columns, which are described in detail as follows. 

Figures~\ref{figure8}a, \ref{figure8}d, and \ref{figure8}g show the reconnection between the flux-rope leg and the ambient sheared arcades, which is also called ar-rf reconnection in the 3D flare model since one post-reconnection field line merges with the flux rope and the other line becomes a flare loop \citep{Aulanier2019}. This reconnection geometry is characterized by the highly curved post-reconnection field lines and significant heating ($\sim 4\;$MK). The $Q$ and $J/B$ distributions denote the formation of the current layer at the interface between the western leg of the flux rope and ambient arcades. One can see that the location of the western footpoint of the flux rope changes after reconnection. Such reconnection is also reflected by the brightening away from the flare ribbons in observations, as shown by the blue box in Figure~\ref{figure4}a. The ar-rf reconnection which likely happens in our simulation changes the connectivity of the field lines and leads to the drifting of the western footpoint of the flux rope \citep{Aulanier2019, Zeman2019}. 

Figures~\ref{figure8}b, \ref{figure8}e, and \ref{figure8}h show the reconnection between the overlying sheared arcades, which is proposed as the standard reconnection paradigm in the 2D CSHKP model, and is called aa-rf reconnection in the 3D flare model \citep{Aulanier2019}. The narrow and vertical current below the flux rope (Figure~\ref{figure8}h) is a proof to demonstrate this reconnection geometry. The ring-like patterns of the QSLs in Figures~\ref{figure8}e reflect the complexity of the flux rope during the eruption, which might be due to the newly-formed twisted field lines and the injected flux. Such QSL structures also reflect that internal reconnection might occur inside the flux rope \citep{Zhong2021}. This type of magnetic reconnection is also manifested in the temperature distributions and the post-reconnection field lines, where the flare loops and the overlying hot cusp-shaped structure revealed in Figure~\ref{figure8}b are the direct evidence for this reconnection geometry \citep{Masuda1994}. Moreover, the border of the CME bubble is heated to 8~MK by the reconnection, which also appeared in other 2D MHD simulations \citep{Mei2012,Zhao2017,Ye2021}.  

Figures~\ref{figure8}c, \ref{figure8}f, and \ref{figure8}i display the reconnection above the flux rope. As the flux rope rises, overlying arcades are pushed outward and reconnect with the ambient antiparallel field lines. This reconnection geometry resembles the breakout model that occurs at the magnetic null above the quadrupole polarities \citep{Antiochos1999}. The reconnected field lines related to this reconnection geometry correspond to the remote ribbon R3 and the southern part of ribbon R1, accounting for the multi-ribbon characteristic of this flare.

\begin{figure}[ht!]
\centering
\includegraphics[scale=0.4]{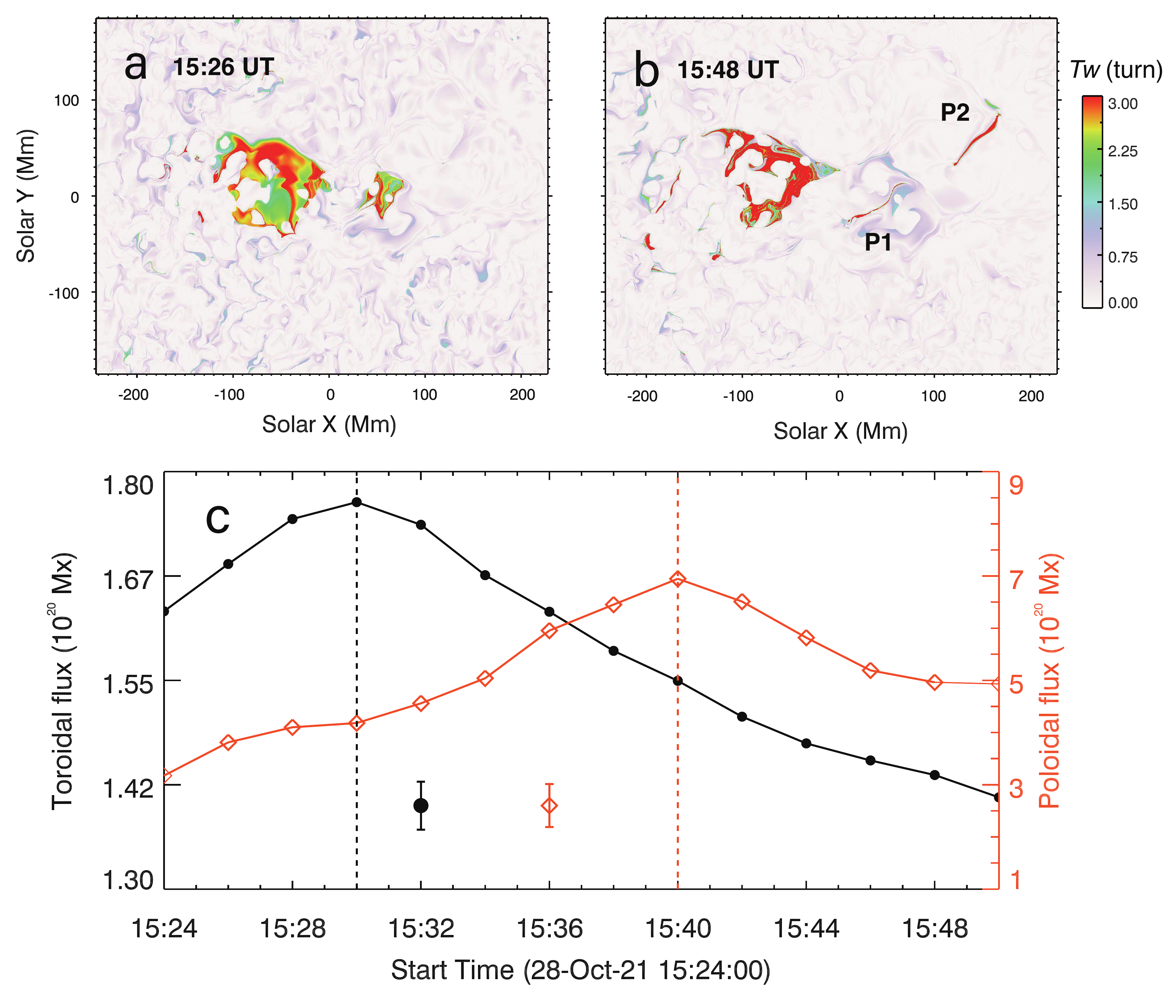}
\caption{Twist distributions on the photosphere at (a) 15:26 UT and (b) 15:48 UT. (c) Temporal evolution of the toroidal (black line) and poloidal (red line) flux of the flux rope. The measurement is repeated 10 times to estimate the errors originating from the uncertainties of the selected regions for the toroidal flux calculation. And the uncertainty of the poloidal flux is calculated by the product of the toroidal flux error and the standard deviation of the twist. The error bars at the bottom represent the mean value of the uncertainty during the whole process. Three reconnection stages are divided by two vertical dashed lines.}
\label{figure9}
\end{figure}

The above-mentioned magnetic reconnection changes the topology of magnetic field, and leads to the variation of the twist ($T_{w}$) inside the flux rope, which is the core structure of the initial magnetic field. In order to  check how magnetic twist changes with time, we calculate the magnetic twist with the parallel electric current method as described in the formula~ (16) in \citet{Berger2006}. According to this method, the initial twist of the flux rope is around 2.16 turns, which is larger than some well-known thresholds of the kink instability \citep{Hood&Priest1981, Torok2004}. In Figures~\ref{figure9}a and \ref{figure9}b we plot two snapshots of the magnetic twist distribution in the photosphere, and it is seen that the twist of the flux-rope footpoint changes a lot during the eruption. First, the twist increases as the reconnection goes on, reaching $\sim$ 2.5 turns at 15:26 UT and $\sim$3 turns at 15:48 UT. Second, the western footpoint splits into two parts, i.e., regions P1 and P2. Region P1 is almost along the polarity inversion line, corresponding to outer newly-formed field lines due to the aa-rf reconnection. Region P2 reflects the drifting of the western footpoint of the flux rope due to the ar-rf reconnection. The distribution of field lines in Figure~\ref{figure3} provides credence to the above physical picture. Figure~\ref{figure9}c shows the temporal evolution of the toroidal ($\phi_{T}=\int \boldsymbol{B_{T}} \cdot \boldsymbol{ds}$) and poloidal fluxes ($\phi_{P}=T_{w} \phi_{T}$), which can be divided into three stages. First, both the toroidal and poloidal fluxes increase, reflecting the aa-rf reconnection, which injects the poloidal and toroidal fluxes into the flux rope. Then, the toroidal flux decreases but the poloidal flux still increases, indicating that the aa-rf reconnection might have stopped and other reconnection geometries begin to work. In particular, the time when the toroidal flux starts to decrease almost corresponds to the time when the flare ribbons cease moving. According to \citet{Aulanier2019}, the toroidal flux should be maintained during the ar-rf reconnection, so we infer that internal reconnection inside the flux rope likely occurs in the second stage. Such a process transfers some of the toroidal flux to the poloidal flux. Finally, both the toroidal and poloidal fluxes decrease, which results from the erosion of the flux rope flux by the null point reconnection above. 

\subsection{Thermal properties of the CME and the flare loops}\label{tp}

\begin{figure}[ht!]
\centering
\includegraphics[scale=0.6]{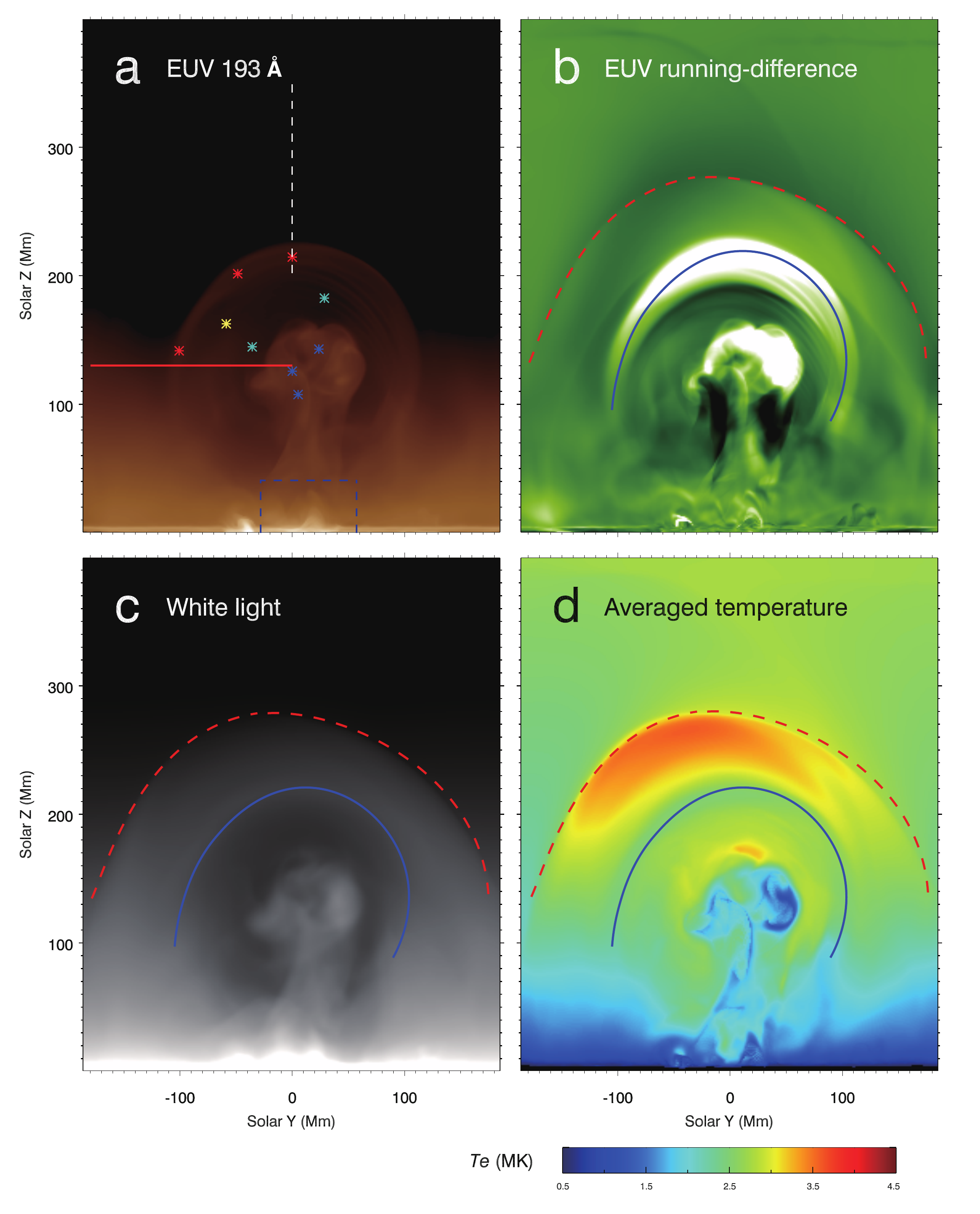}
\caption{(a) Synthesized EUV 193 \AA\ image at 15:32 UT. (b) Running-difference image of the EUV 193 \AA\ radiation at 15:32 UT, where the blue solid line outlines the CME leading edge, and the red dashed line outlines the piston-driven shock. (c) Synthesized white-light image at 15:32 UT. (d) Averaged temperature distribution at 15:32 UT.}
\label{figure10}
\end{figure}

Figure~\ref{figure10}a displays the synthesized SDO/AIA 193~\AA\ emission image viewed from the $x$-axis (along the flux-rope axis) at 15:32 UT. It is seen that the EUV image presents the three-component structures of CMEs, i.e., a bright front, a cavity, and bright core, with the solar flare below the bright core near the solar surface. In order to show the structures more clearly, we make a corresponding running-difference image in Figure~\ref{figure10}b, where the 193 \AA\ image at 15:30 UT is subtracted. Besides the three components of a typical CME (where the frontal edge is indicated by the blue solid line), a faint dome-like front is identified above the EUV frontal front as indicated by the red dashed line, which should correspond to the piston-driven shock. In order to compare our simulation results with coronagraph observations, we synthesize the white-light image, which is shown in Figure~\ref{figure10}c. The Thomson scattering of the white-light intensity is calculated with the eltheory.pro in the Solar SoftWare, which considers the white-light intensity to be proportional to the integration of the density along the line of sight. As shown in Figure~\ref{figure10}c, the white-light image of the eruption is characterized by the typical three-component structures, i.e., a leading front, a dark cavity, and a bright core. However, it should be mentioned that the leading front is comprised of two parts, an inner part as indicated by the blue solid line which corresponds to the EUV frontal edge, and an outer part as indicated by the red dashed line which corresponds to the suspected piston-driven shock. There is no distinct gap between the two parts although the intensity in between is slightly weaker than the leading and trailing fronts. To understand the thermal properties of the suspected shock and the three components of the CME, we calculate the averaged temperature of each pixel in Figure \ref{figure10}c by $\bar{T}=\int n_{H}^{2} Tdx$/$\int n_{H}^{2} dx$ \citep{Cheng2012, Gou2015}, and plot the $\bar{T}$ distribution in Figure~\ref{figure10}d. It is seen that the suspected shock, as indicated by the red dashed line, corresponds to a drastic increase of temperature, and the downstream of the suspected shock is strongly heated, which is in accordance with the shock theory. In order to confirm the shock nature of this front, we calculate the propagation velocity at the nose part based on the synthesized images at 15:32 UT. It is found that the propagation speed is 590.2 km s$^{-1}$, which is larger than the local fast-mode MHD wave speed with 268.3 km s$^{-1}$. Our result confirms that the dome-like structure as indicated by the red dashed line in Figure \ref{figure10}b is indeed a fast-mode shock wave with a fast-mode Mach number of 2.1. It is interesting to see that while the EUV bright edge, as indicated by the blue solid line, is only slightly heated, the cavity is moderately heated. The CME bright core has a dichotomous distribution of temperature, with the central part being cold and the shell part being warm. 

\begin{figure}[ht!]
\centering
\includegraphics[scale=0.2]{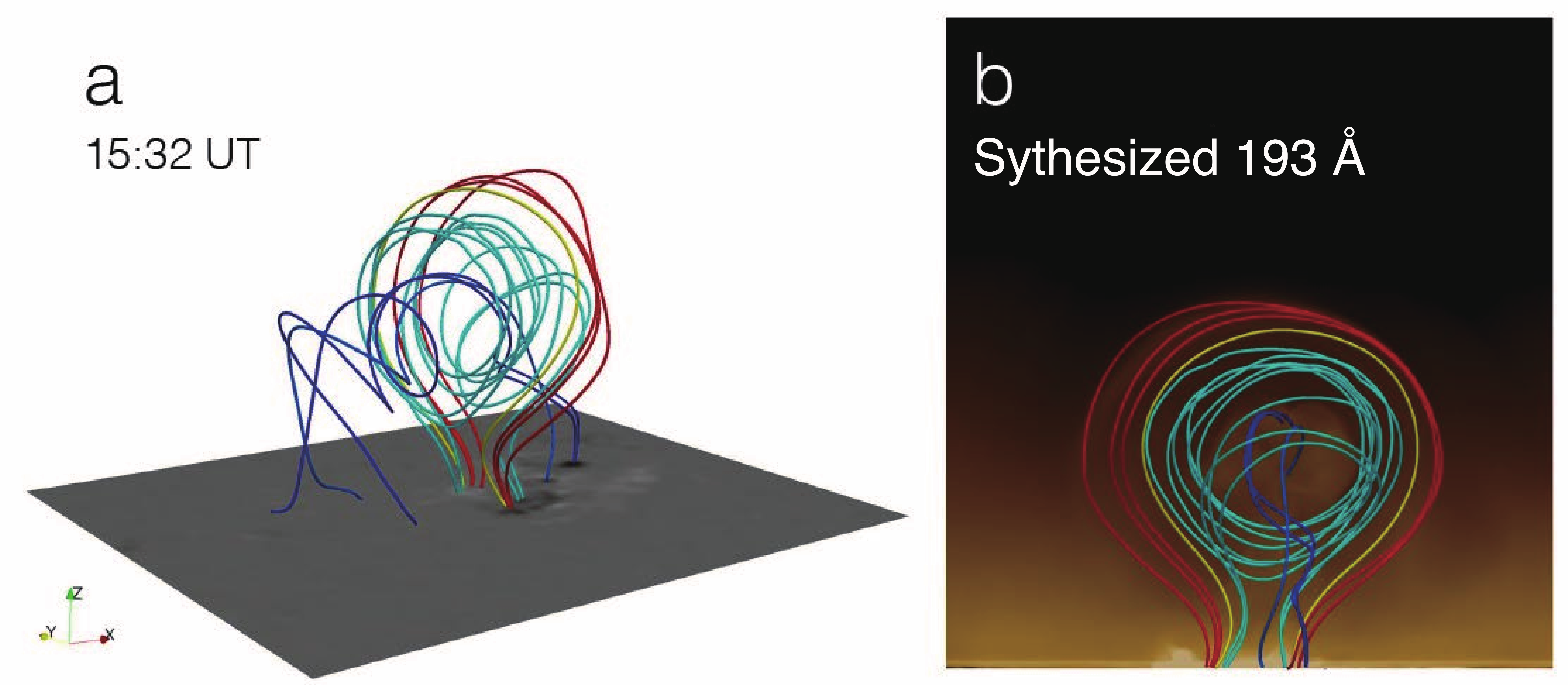}
\caption{Field lines passing through the asterisks in Figure~\ref{figure10}a (red: CME leading edge; cyan and yellow: CME cavity; blue: CME core) on the plane of $x=$10 Mm. Panel (a) shows the side view, and panel (b) shows the end view overlaid on the 193 \AA\ image.}
\label{figure11}
\end{figure}

The corresponding magnetic field lines at 15:32 UT are displayed in the left panel of Figure \ref{figure11}, where the strongly twisted core field and the less twisted envelope field can be clearly distinguished. The magnetic field geometry viewed from the same vantage point as Figure \ref{figure10} are displayed in the right panel of Figure~\ref{figure11}, with the background being the synthesized 193 \AA\ intensity map. One can see that the bright core, wrapped by twisted field lines, occupies slightly more than the lower half of the flux rope. As for the dark cavity, we find that the plasma therein is hot and tenuous, and the magnetic field lines consist of two portions, inner and outer portions: the inner field lines are hot and twisted (cyan lines in Figure~\ref{figure11}), but the outer portion is embedded by non-twisted magnetic loops (yellow lines in Figure~\ref{figure11}). It implies that part of the cavity is due to the upper portion of the twisted flux rope with low density, and another part is due to magnetic field-line stretching as described by \citet{Chen2011}. Regarding the EUV bright front (which is also the CME frontal edge), the plasma there is moderately heated (2~MK) and compressed (blue solid lines in Figure~\ref{figure10}), corresponding to the outer edge of the stretching field lines (red lines in Figure~\ref{figure11}) according to the magnetic field-line stretching model \citep{Chen2002}, which is the typical characteristics of the EUV wave \citep{Liuw2014}. 

\begin{figure}[ht!]
\centering
\includegraphics[scale=0.8]{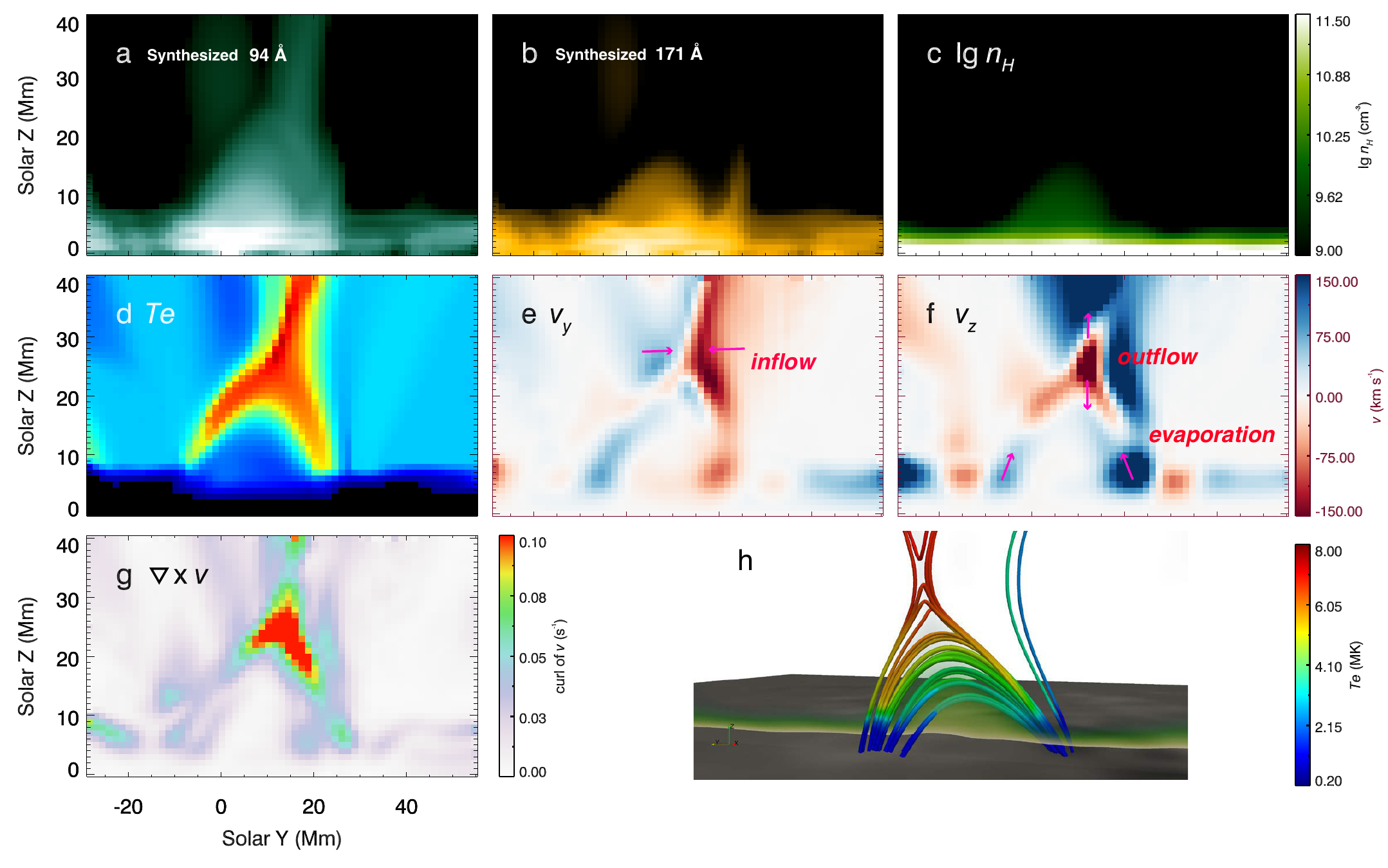}
\caption{Panels (a) and (b) respectively show the synthesized 94 \AA\ and 171 \AA\ images integrated from the $x=7$ Mm to 21 Mm zoomed in the rectangle region in Figure~\ref{figure10} a. (c) Number density distribution on the plane of $x=$14 Mm. (d) Temperature distribution on the plane of $x=$14 Mm. (e) $v_{y}$ distribution on the plane of $x=$14 Mm. (f) $v_{z}$ distribution on the plane of $x=$14 Mm. (g) The curl of the velocity distribution on the plane of $x=14$ Mm. (h) Magnetic field lines colored in temperature near the cusp-shaped structure, where the background transparent slice represents the number density distribution in panel (c).}
\label{figure12}
\end{figure}

A cusp-shaped structure is widely perceived as important evidence for magnetic reconnection \citep{shib99}, which is worthy of exploration in a data-driven simulation. Figures~\ref{figure12}a and \ref{figure12}b display the synthesized 94~\AA\ and 171~\AA\ images, respectively. It is seen that the cusp-shaped structure can only appear in high-temperature waveband (94~\AA, $\sim 10\;$MK ), and post-flare loops are more visible in the low-temperature 171 \AA\ image ($\sim 0.8\;$MK). Figures~\ref{figure12}c and \ref{figure12}d depict the number density and temperature distributions, respectively. We find that the cusp-shaped structure is hot (8 MK), whereas the post-flare loops are relatively cold ($< 1\;$MK) and dense, as expected. Figures~\ref{figure12}e and \ref{figure12}f show the velocity distributions, where we can recognize the inflow and outflow near the reconnection site, and the chromospheric evaporation at the loop footpoints. Figure~\ref{figure12}g illustrates the curl of the velocity, which is generally used to detect the slow-mode shock \citep{Wang2009, Mei2020}. It is seen that the slow-mode shock is located near the hot reconnection site, confirming that it plays a significant role in heating the reconnection outflow as claimed in the reconnection model \citep{Priest1986}. Figure~\ref{figure12}h illustrates the field lines color-coded in temperature. Clearly, one can see the highly curved and hot cusp-like structures, as well as cold and rounded flare loops, which is similar to the scenario proposed by \citet{shib99}. 

\section{Discussions}\label{sec:discussion}

\subsection{Stopping positions of flare ribbons}

Magnetic topology governs the global dynamics of solar fares \citep{long01}. According to the magnetic reconnection model \citep{Priest1986}, magnetic energy is released to produce heat and non-thermal particles around the reconnection site, which are then transferred to the solar surface along magnetic separatrices. As a result, two flare ribbons usually appear at the bases of the separatrices, which run roughly parallel with the polarity inversion line. As time goes on, the flare loops become bigger and bigger apparently and the two flare ribbons separate. The moving speed of each ribbon can be more than 50 km s$^{-1}$ in the impulsive phase and then decreases to even less than 1 km s$^{-1}$. It is noted in passing that the separating motion of the flare ribbons was frequently attributed to the rising reconnection point. However, as pointed out by \citet{Chen1999}, this is a misinterpretation, and the flare ribbons would separate even if the reconnection point is fixed once the reconnection point is high enough.

There were a variety of efforts revealing the cospatiality between flare ribbons and separatrices \citep{somo02, demo93, Dalmasse2015, Zuccarello2017}, or in more general cases, QSLs \citep{demo96, Titov2002}, which confirmed the validity of the magnetic reconnection model for solar flares. In all these works, the separatrices or QSLs were calculated from the magnetic field extrapolation derived from the photospheric magnetograms. \citet{Chen2012} argued that the flare ribbon-associated  QSLs are dynamic, whereas the QSLs derived from the static extrapolated coronal field are fixed. Therefore, they proposed that there should exist two types of QSLs, i.e., inner and outer QSLs. The inner QSLs are linked to the reconnection site, and are cospatial with the moving flare ribbons, whereas the outer QSLs correspond to the location where the flare ribbon would eventually stop. This argument was reinforced by the fact that the extrapolated QSLs are cospatial with the flare ribbons at their full extents \citep{savc15}.

To simulate the dynamic QSLs, \citet{ savc16} used the magnetofrictional flux-rope insertion method to generate an unstable coronal magnetic field, where the iteration of the coronal magnetic field relaxation leads to continuous reconnection under the flux rope rather than approaching an equilibrium state. Although the relaxation is not equivalent to dynamic evolution, it indeed captures the topological variations of CME/flare eruptions, including the expansion of flaring loops and the separation of flare ribbons. Of course, a better approach to compare dynamic QSLs and flare ribbons is to conduct MHD numerical simulations \citep{Jiang2018a, Guo2019a, Zhong2021}. While quantitative comparisons were made by several research groups, \citet{Jiang2018a} performed the first quantitative comparison between separation speeds of the simulated QSLs and observed flare ribbons. However, their simulated QSLs separate with a speed three times faster than that of observed flare ribbons, which they attributed to a faster reconnection rate due to high numerical resistivity in their simulations. 

In this paper, we performed a data-driven MHD simulation in order to reproduce the CME/flare eruption event on 2021 Oct 28, and then compared the simulated QSLs location with the observed flare ribbons. As shown in Figure~\ref{figure4}e, the inner QSLs, which are linked to the reconnection site, and almost cospatial with the observed flare ribbons during the whole eruption. The simulated separating speeds of the two QSLs in the early stage are 120 km s$^{-1}$ and 19 km s$^{-1}$, respectively, both of which are consistent with the observed separating speeds of the two flare ribbons. More importantly, we found that both the simulated QSLs and observed flare ribbons eventually stop. As seen from Figure~\ref{figure4}e, the locations where the simulated QSLs and the observed flare ribbons stop are at the pre-existing QSLs, which correspond to the outer QSLs as defined by \citet{Chen2012}.

\begin{figure}[ht!]
\centering
\includegraphics[scale=1.6]{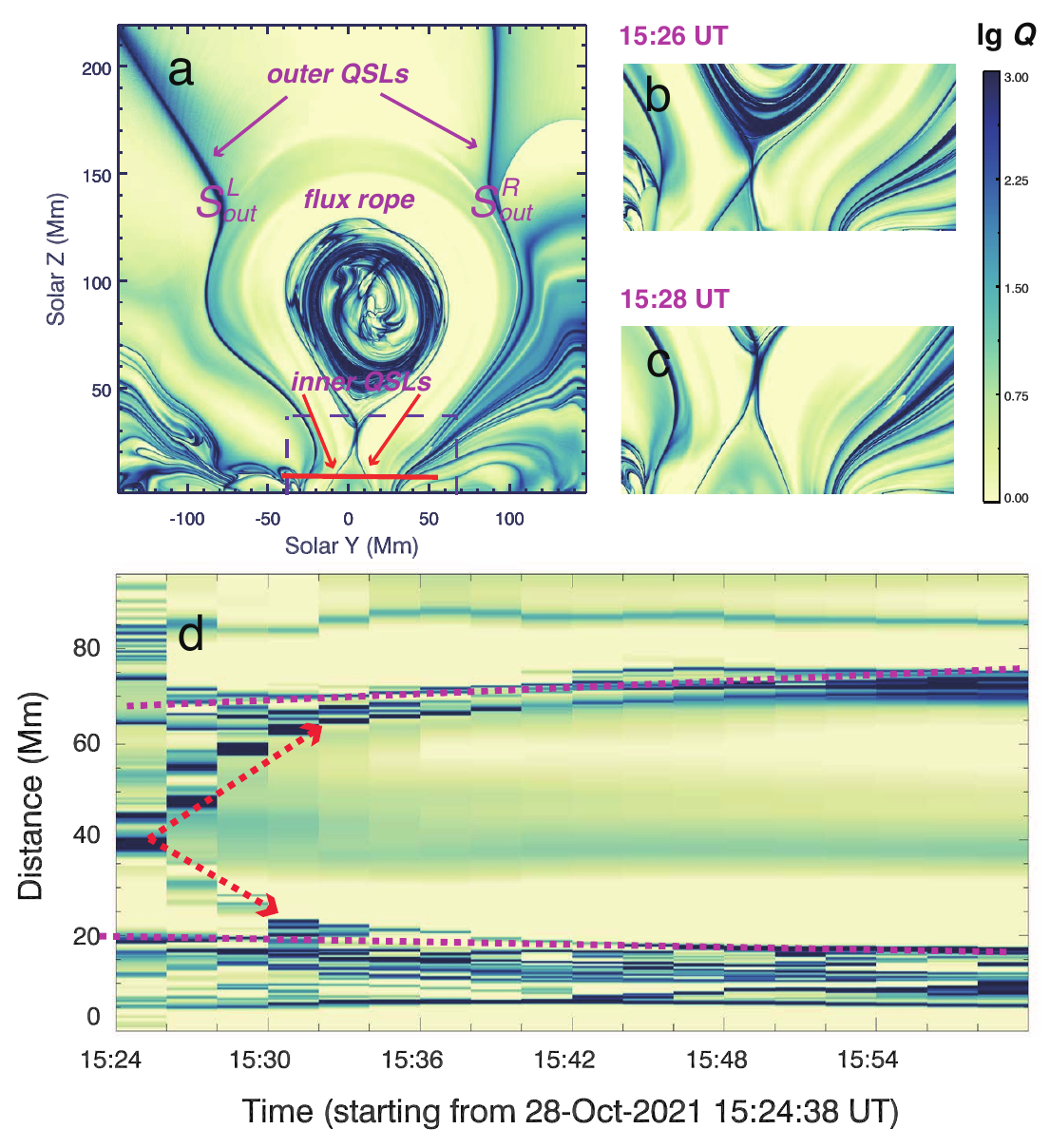}
\caption{(a) $Q$ distribution on the plane of $x=$10.6 Mm at 15:28 UT. Panels (b) and (c) illustrate the Q distributions zoomed in the purple rectangle region in panel (a) at 15:26 and 15:28 UT, respectively. (d) Time-distance diagram of $Q$ along the slice  (red solid line) in panel (a).}
\label{figure13}
\end{figure}

To illustrate the relationship between the inner separating QSLs and the almost fixed outer QSLs, in Figure~\ref{figure13}a we plot the distribution of the squashing factor $Q$ in the cross-section of the simulation domain at $x=10.6$ Mm at 15:28 UT. The cross-section is in the $y$--$z$ plane, which is roughly perpendicular to the axis of the flux rope. One can see an erupting flux rope in the middle, with the outer shell extending down to form an X-shaped hyperbolic flux tube, which corresponds to the inner QSLs, as indicated by the red arrows. The inner QSL further extends down, bifurcating into two branches, and their intersection with the bottom boundary corresponds to the two separating flare ribbons. Meanwhile, we can see in Figure~\ref{figure13}a that beyond the left and right sides of the flux rope, there stand two QSLs, as indicated by the pink arrows, which are labeled as $S_{\rm out}^{L}$ for the left QSL and $S_{\rm out}^{R}$ for the right QSL. Only the magnetic field lines between the two outer QSLs and the two inner QSLs are available for further reconnection. Hence, as the dynamic inner QSLs move outward and approach the outer QSLs, magnetic reconnection stops. As seen from  Figures~\ref{figure13}b and \ref{figure13}c, the inner QSLs move outward as the reconnection goes on, while the outer QSLs are almost fixed in the photosphere. Moreover, the time-distance diagram shown in Figure~\ref{figure13}d illustrates that the dynamic inner QSLs eventually stop at the static outer QSLs, as expected. It is noticed that in our simulation the time when the two flare ribbons stop is very close to the moment when the toroidal flux starts to decrease, which implies that the outer QSLs determine where the aa-rf reconnection stops.

We perceive that this property is profound and might have various applications in future space weather prediction. First, the co-spatiality between QSLs and flare ribbons is an effective touchstone to examine the rationality of any data-driven simulations. Second, the outer QSLs, which can be determined by the magnetograms before eruption happens, determine directly the size of the flare ribbons and reconnection regions. It means that we can estimate the magnetic flux to be injected into the CME due to reconnection even before the flare, and then infer the total magnetic flux and the twist of an ICME \citep{Qiu2007, Wangws2017, Xing2020b}. Moreover, with some assumptions, we can even estimate the lifetime of a flare before it occurs. Consequently, we strongly suggest that the outer QSLs and the above-mentioned properties can be considered as novel indexes of future space weather forecasting.

\subsection{Relationship between CME fronts and EUV waves}

A typical CME is generally characterized by a frontal edge, with a cavity and bright core embedded inside. When a CME is fast enough, a piston-driven shock would straddle over the CME bubble \citep{Chen2011}, as revealed in observations \citep{vour03}. Albeit this phenomenon has been observed for several decades, the consensus as to its nature has not been obtained hitherto. The reason is that it involves numerous elaborated physical processes, in particular the thermodynamics, some of which are omitted in the traditional zero-$\beta$ or isothermal MHD models. Among the portions of CME three-part structures, the most controversial one is the bright leading front. Its nature and formation mechanism are still elusive.

In earlier times, the CME frontal edge was attributed to be MHD waves or erupting coronal loops \citep{Nakagwa1975, Poland1976}. Later, it has been assumed to be plasma pileup ahead of the erupting flux rope \citep[see][for a review]{forb00}. However, \cite{Chen2009} proposed a conjecture to explain it with the field-line stretching model \citep{Chen2002}. He investigated the relationship between ``EIT wave'' and CME leading front by comparing the white-light coronagraph images in the high corona and low-corona EUV observational data, where the ``EIT wave" corresponds to the slow component of the coexisting two EUV waves. It is worth noting that, as suggested by \citet{Chen2016}, we are inclined to use ``EUV waves" for any kind of wavelike phenomena observed in EUV wavebands, including fast and slow components in this paper, and use ``EIT waves" for the wavelike phenomena discovered by \citet{Thompson1998} with the SOHO/EIT telescope, which are typically three times slower than type II radio bursts and Moreton waves \citep{Klassen2000}, i.e., ``EIT waves" correspond to the slow-component EUV waves in this paper. It was found that the CME front edge is almost cospatial with the ``EIT wave'', and the following dimming region corresponds to the CME dark cavity. Therefore, he applied the formation mechanism of ``EIT waves". i.e., magnetic field-line stretching model, to the CME frontal edge:  As a flux rope rises, it pushes the overlying field lines to stretch successively, producing the density disturbances that propagate upward with the fast-mode wave velocity. Simultaneously, such disturbances are also transferred down to the footpoints along field lines with the Alfv$\acute{\rm e}$n velocity. The density enhancements at different portions of field lines at a given time compose the CME front. The enclosing volume increases as the field lines are stretched, forming the coronal dimming and the CME cavity. As such, the ``EIT wave'' and the following dimming correspond to the CME leading front and dark cavity, respectively \citep{Chen2009}.

\begin{figure}[ht!]
\centering
\includegraphics[scale=0.8]{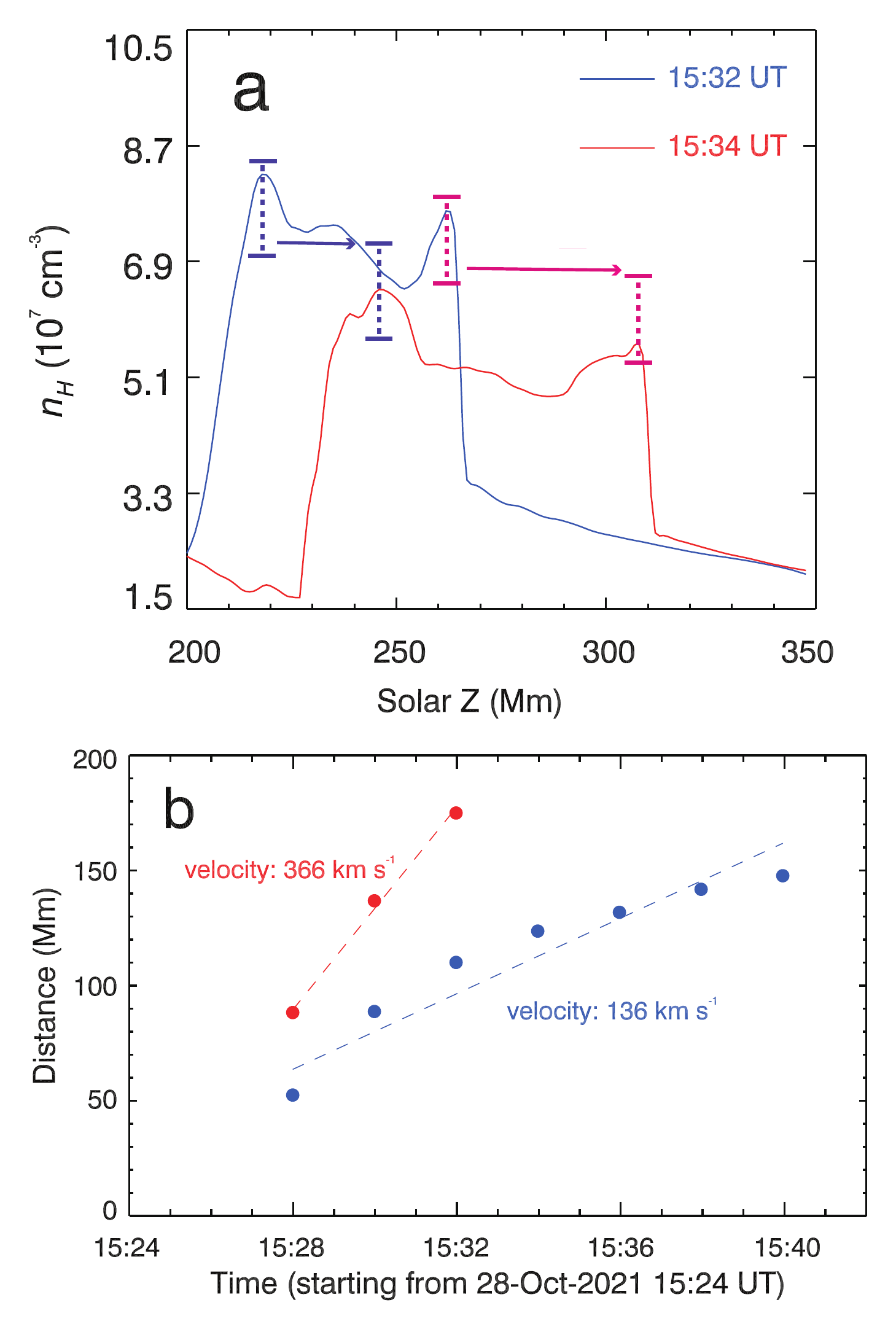}
\caption{Panels (a) illustrates the number density distributions at 15:32 UT and 15:34 UT along the white dashed line in Figure~\ref{figure10}a, respectively. Panel (b) illustrates the time-distance diagram of two wave-like structures along the red solid line in Figure~\ref{figure10}a.}
\label{figure14}
\end{figure}

The above theory is verified very well by our data-driven model. As shown in Figures~\ref{figure10}b and \ref{figure10}c, the synthesized EUV images based on our simulation reproduce two components of EUV waves and the ensuing EUV dimmings. In particular, it is confirmed that the dimmings and the slow-component EUV wave correspond to the synthesized CME cavity and CME leading front, respectively, whereas the fast-component EUV wave corresponds to the CME piston-driven shock. To further understand the physical nature of the EUV waves, we plot the density distributions at two moments in Figures~\ref{figure14}a along the white dashed line in Figure~\ref{figure10}a. There are two fronts with enhanced density, i.e., a leading front which travels faster and a following front which travels slower. Besides, a dimming region is found to be immediately after the slower front. All these features are analogous to the MHD simulation performed by \citet{Chen2002}. Then, we calculate the velocities of these wave-like structures and local fast-mode waves. We find that the leading fast front moves faster than that of the local fast-mode wave ($\sim 2.05\ V_{\rm fw}$), but the following slow front almost propagates at the local fast-mode wave ($\sim 0.86\ V_{\rm fw}$). As mentioned in \citet{Chen2009}, a rising flux rope can produce two fronts propagating upward, i.e., a piston-driven shock front and a following dome-shaped front. The latter corresponds to the slow-component EUV waves and the CME leading front, where the top part travels upward with the fast-mode wave speed and the legs travel horizontally with a speed about three times slower than the fast-mode wave speed \citep{Chen2002, Chen2005}.

Since EUV waves are more frequently observed across the solar disk, we examine the time-distance evolution of these two waves in the horizontal direction. The horizontal slice is marked as the red solid line in Figure~\ref{figure10}a. The locations of the legs of the two EUV waves at several moments are displayed in Figure~\ref{figure14}b. It is found that the velocity of the fast-component EUV wave is about 2.69 times that of the slow component. Consequently, we believe that the two components EUV waves in our simulation, i.e., the fast and slow components, correspond to the fast piston-driven shock and the non-wave component predicted by the field-line stretching model \citep{Chen2002, Chen2005}. It is noticed that while the fast-component wave decelerates slightly, the slow-component wave decelerates significantly. According to the magnetic stretching model \citep{Chen2002, Chen2005}, the apparent speed of the slow-component EUV wave is not only related to the magnetic field strength, but also to the magnetic configuration. If the magnetic field lines become more and more elongated in the vertical direction, as indicated by Fig. 8 of \citet{Chen2005}, the slow-component EUV wave would decelerate substantially. The slow-component EUV wave is three times slower than the fast-component EUV wave only when the magnetic field lines are concentric semi-circles \citep{Chen2002}. It is noted that recent observational analysis for this event performed by \citet{Devi2022} distinguished two components of EUV waves in this event, including the leading fast-mode wave and the following slow non-wave component, verifying the field-line stretching model of EUV waves. Moreover, they found that the ratio between the speeds of these two components ranges from 2.0 to 2.5, which is fairly closer to our simulation result. Therefore, our data-driven simulation provides a straightforward verification for the field-line stretching model \citep{Chen2002, Chen2005}, and confirms that the slow-component EUV wave is cospatial with the CME frontal edge \citep{Chen2009}.

\begin{figure}[ht!]
\centering
\includegraphics[scale=0.5]{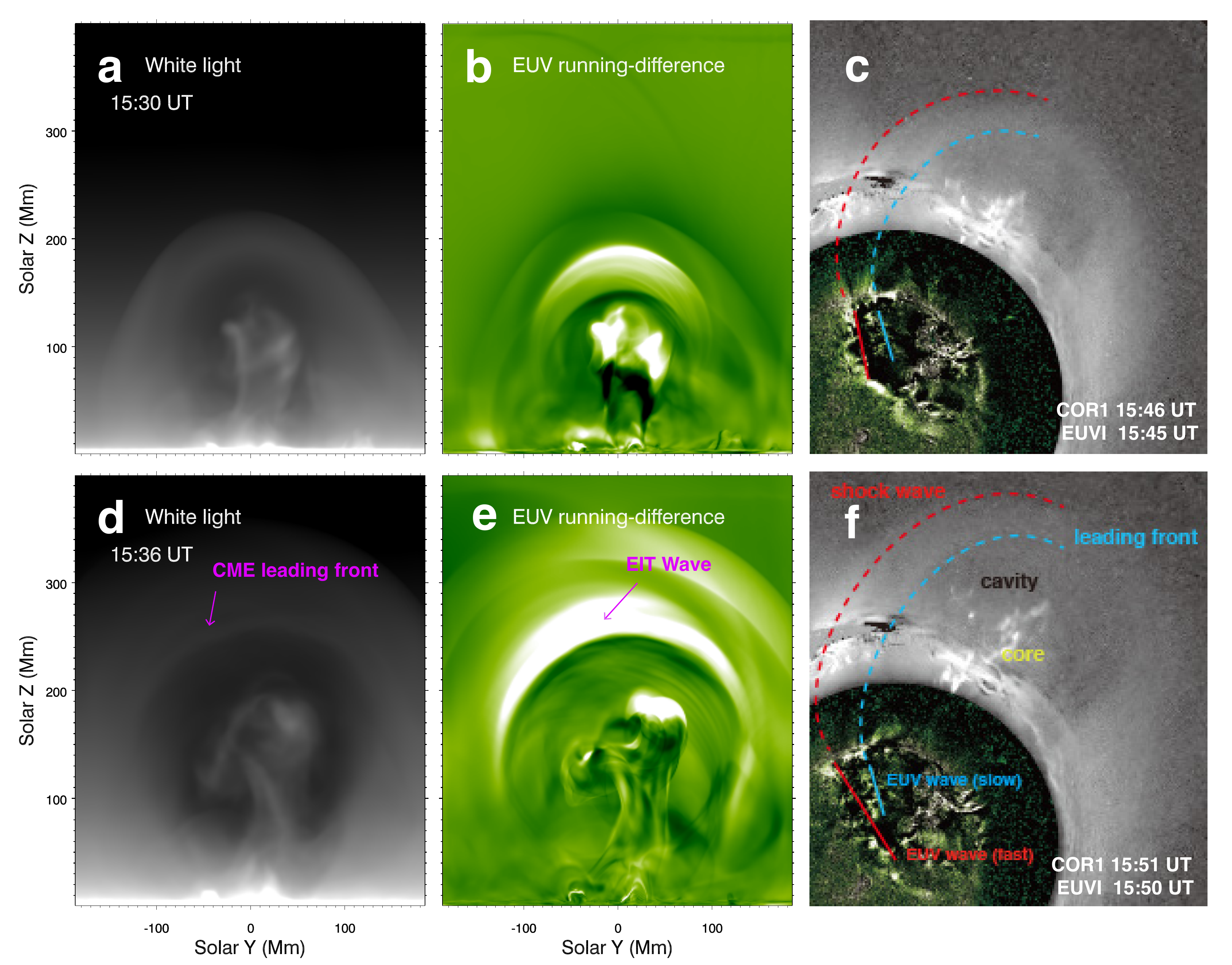}
\caption{Panels (a) and (d) respectively show the side views of the synthesized white-light images at 15:30 UT and 15:36 UT. Panels (b) and (e) respectively show the side views of the synthesized EUV 193 \AA\ running-difference images at 15:30 UT and 15:36 UT. The pink arrows in panels (d) and (e) indicate the CME leading front and ``EIT wave'', respectively. Panels (c) and (f) show the STEREO/COR1 images with the superposition of EUVI 195 \AA\ running-difference images at 15:46 and 15:51 UT, respectively. The blue (red) dashed lines outline the CME leading front (forward shock wave), and the solid lines outline the two components of EUV waves.}
\label{figure15}
\end{figure}

To explore the nature of the CME frontal edge, as done in observations \citep{Chen2009}, we compare the EUV waves in the synthesized EUV images and three-part structures of the CME in the synthesized white-light image in Figure~\ref{figure15}. One can see that the fast- and slow-component EUV waves are almost cospatial with the shock and the CME frontal edge, respectively, and the EUV dimmings are cospatial with the CME cavity as well. To compare the synthesized images with observations, we plot the composed COR1 white-light and EUVI 195 \AA\ images observed by STEREO A at two moments in Figures~\ref{figure15}c and \ref{figure15}f, where the COR1 and EUVI observations are roughly simultaneous. Owing to the limited field of view of the EUVI, we cannot compare the synthesized EUV images with EUVI observations directly. However, we can compare the synthesized white-light images with the COR1 observations. As revealed in the panels a, c, d, and f of Figure~\ref{figure15}, the CME frontal edge and the piston-driven shock resemble the observations very well. Generally it is often taken for granted that the CME piston-driven shock should be distinguished clearly from the CME frontal edge as seen in some events. However, it is noted that in other events like this one, the CME frontal edge and the piston-driven shock above become inseparable because the density in the downstream (or the sheath) region of the piston-driven shock is also enhanced, and the CME frontal edge and the CME shock join together as a unity as revealed in Figures~\ref{figure15}c and \ref{figure15}f. This result should be kept in mind when identifying the CME frontal edges and piston-driven shock waves in some CME events.

Therefore, our simulation results strongly indicate that the CME frontal edge corresponds to the density enhancement caused by the successive stretching of the erupting magnetic field lines rather than the plasma pileup \citep{Chen2009}. That is to say, the CME frontal edge is the same as ``EIT wave'' in the formation mechanism and physical nature, i.e., both are due to successive stretching of magnetic loops.

\section{Summary}\label{sec:summary}

In this paper, we perform a data-driven radiative MHD simulation to study the thermodynamic and magnetic topology evolutions of the X1.0 flare on 2021 October 28. We adopt a non-adiabatic MHD model, including the thermal conduction and radiative losses. The initial magnetic field with a flux rope is provided by the RBSL technique, which is then relaxed to a force-free state with the MF method. The evolution is driven by the changing bottom boundary conditions, including a sequence of vector magnetic field and velocity field obtained from the observations. Our simulation reproduces many observational features in the corona and sheds light on some underlying physical processes, which are summarized as follows. 
\begin{enumerate}
\item{Our simulation results are comparable with observations in many sences. First, the simulated magnetic flux rope resembles the filament, and possesses the analogous eruption direction to the observations. Second, the simulated QSLs in the photosphere are almost cospatial with the observed flare ribbons. Particularly, the typical separation motion of flare ribbons is reproduced very well by our numerical model. Third, the synthesized EUV images reproduce some general observational features in CME/flare eruptions, for example, the gross morphology of the flare, helical prominence, three-part CME structure, and flaring loops. Given all of these, we believe that our data-driven simulation has reproduced the principal observational features and physical processes, encompassing the thermodynamic and magnetic topology evolutions.}

\item{With the aid of the field-line connectivity, heating, QSLs, $J/B$, and flux evolution, we identified three main types of reconnection geometries during the eruption in our simulation, including the reconnection in the overlying arcades, the reconnection between flux-rope legs and adjacent arcades, and the breakout-like reconnection above the flux rope. Such reconnection geometries greatly affect the topology of the flux rope and eruption dynamics. Our data-driven radiative model strongly suggests that the reconnection in real observations are far more complicated than described in the 2D CSHKP model.}

\item{Our numerical model provides conclusive validation for the conjecture that the separating flare ribbons, which correspond to the dynamic QSL,  ultimately stop at the outer stationary QSLs \citep{Chen2012}. The outer QSLs represent the interface beyond which magnetic field is not involved in magnetic reconnection associated with the flare. When all field lines inside the outer QSLs have reconnected, the flare ribbons would stop at this separatrix. This finding provides a new window to predict the size of the flare and reconnection regions before eruption, which could be used in future space weather forecasting.}

\item{Our numerical model reveals the cospatiality between CME piston-driven shock and the fast-component EUV wave, as well as the cospatiality between the CME leading front and ``EIT wave'', i.e., the slow-component EUV wave. The simulation reproduces the CME leading front followed by a dark cavity, which correspond to the ``EIT wave'' and the dimming region, respectively. Besides, the ``EIT wave'' speed is about a third of the piston-driven shock ahead when propagating along the solar surface, supporting the magnetic field-line stretching model for the ``EIT waves'' \citep{Chen2002}.}
\end{enumerate}

It is worth pointing out that this event has also been studied by \citet{Yamasaki2022} with their zero-$\beta$ data-constrained simulation, focusing on the eruption mechanism and the reconnection process. Some differences emanate from the comparison between these two models. In their paper, they constructed the initial NLFFF with the direct extrapolation, so the eastern portion of the filament located above weak fields fails to be constructed very well. Accordingly, the flare ribbons in this region are barely reproduced. In our paper, we construct the initial magnetic field with the RBSL flux-rope technique and MF relaxation, which is particularly capable to construct the flux rope that is in agreement with the observational filament located in weak-field regions. As a result, the corresponding flare ribbons in our simulations are more accordant with observed ribbons. Therefore, we conclude that the RBSL flux-rope model is extremely advantageous to construct the twisted magnetic structure and further study the eruption originating from the weak-field regions.

Although our simulation has reproduced many observational features and revealed several physical processes, the deficiencies are also apparent. First,  flare ribbon R2 is not evident in the synthesized EUV images. We perceive that the primary reason is probably the resolution of our simulation. Owing to the restrictions of the computing resources and observational resolutions, the grid spacing in our simulation is about 1 Mm, which is significantly lower than in other numerical radiative MHD models. For example, the highest spatial resolution can attain 192 km in \citet{Cheung2019}, 64 km in \citet{Chenf2022} and 25 km in \citet{Chenyj2021}. Thus, the energy exchange between the corona and transition region might be difficult to be captured very well in our simulation. The second reason might be the treatment of energy terms. Our simulation adopts the optically thin radiative losses and ignores the radiation transfer in the low atmosphere, which is likely to affect the morphology of the flare ribbons. Other than these, our MHD model neglects the contribution of the non-thermal energetic particles, which enhance the chromsopheric evaporation and the brightening of flare ribbons \citep{Kontar2011, Ruan2020}. The contribution of fast electrons should be more significant for flare ribbon R2 due to the magnetic mirror effect \citep{Yangyh2012}, since it has weaker magnetic field than flare ribbon R1. We believe that future data-driven radiative MHD simulations combined with fast electron physics and radiation transfer are bound to better match the observations. In addition to the mismatch of flare ribbon R2, the second drawback of our simulation is the driving duration. This simulation only involves the eruption phase of the flux rope, whereas the buildup process of the magnetic free energy is not simulated, being substituted by the NLFFF model. In future work, we attempt to develop the long-term evolution of the active region and pay more attention to the buildup process of the CME source region.

\acknowledgments
The numerical calculations in this paper were performed in the cluster system of the High Performance Computing Center (HPCC) of Nanjing University. The SDO data are available courtesy of NASA/SDO and the AIA and HMI science teams. The EUVI data are available courtesy of the STEREO/SECCHI consortium. This research was supported by National Key Research and Development Program of China (2020YFC2201200, 2022YFF0503004), NSFC (12127901, 11961131002, and 11773016, 11533005). J.H.G was supported by China Scholarship Council under file No. 202206190140. SP acknowledges support from the projects C14/19/089  (C1 project Internal Funds KU Leuven), G.0D07.19N  (FWO-Vlaanderen), SIDC Data Exploitation (ESA Prodex-12), and Belspo project B2/191/P1/SWiM.

\bibliography{ms}{}
\bibliographystyle{aasjournal}

\newpage

\end{document}